\documentclass{aa} 
\usepackage{graphicx}
\usepackage{txfonts}
\usepackage[usenames]{color} 
\usepackage{tabularx} 
\usepackage{footnote}
\makesavenoteenv{table*}
\usepackage{scalerel}
\usepackage{tikz}
\usetikzlibrary{svg.path}
\definecolor{orcidlogocol}{HTML}{A6CE39}
\tikzset{
  orcidlogo/.pic={
    \fill[orcidlogocol] svg{M256,128c0,70.7-57.3,128-128,128C57.3,256,0,198.7,0,128C0,57.3,57.3,0,128,0C198.7,0,256,57.3,256,128z};
    \fill[white] svg{M86.3,186.2H70.9V79.1h15.4v48.4V186.2z}
                 svg{M108.9,79.1h41.6c39.6,0,57,28.3,57,53.6c0,27.5-21.5,53.6-56.8,53.6h-41.8V79.1z M124.3,172.4h24.5c34.9,0,42.9-26.5,42.9-39.7c0-21.5-13.7-39.7-43.7-39.7h-23.7V172.4z}
                 svg{M88.7,56.8c0,5.5-4.5,10.1-10.1,10.1c-5.6,0-10.1-4.6-10.1-10.1c0-5.6,4.5-10.1,10.1-10.1C84.2,46.7,88.7,51.3,88.7,56.8z};
  }
}

\newcommand\orcid[1]{\href{https://orcid.org/#1}{\mbox{\scalerel*{
\begin{tikzpicture}[yscale=-1,transform shape]
\pic{orcidlogo};
\end{tikzpicture}
}{|}}} \href{#1}{#1}}
\usepackage[breaklinks=true,colorlinks, linkcolor=blue,urlcolor=blue,citecolor=blue]{hyperref}

\begin{document}

   \title{The unusual widespread solar energetic particle event on 2013 August 19}

   \subtitle{Solar origin and particle longitudinal distribution}

   \author{L. Rodríguez-García
          \inst{1}
          \and
          R. Gómez-Herrero\inst{1}
          \and I. Zouganelis\inst{2} \and L. Balmaceda\inst{3,4} \and T. Nieves-Chinchilla\inst{3} \and \newline  N. Dresing\inst{5,6} \and 
          M. Dumbovi\'c\inst{7} \and N. V. Nitta\inst{8} \and F. Carcaboso\inst{1}  \and L. F. G. dos Santos\inst{9}\and L. K. Jian\inst{3} \and \newline L. Mays\inst{3} \and D. Williams\inst{2}  \and  J. Rodríguez-Pacheco\inst{1}
          }
   \institute{Universidad de Alcalá, Space Research Group, Alcalá de Henares, Madrid, Spain \\
              \email{l.rodriguezgarcia@edu.uah.es}
         \and
             European Space Astronomy Center, European Space Agency, Villanueva de la Cañada, Madrid, Spain
             \and
            Heliophysics Science Division, NASA Goddard Space Flight Center, Greenbelt, MD, USA
             \and
             George Mason University, Fairfax, VA, USA
             \and
            Institut fuer Experimentelle und Angewandte Physik, University of Kiel, Kiel, Germany
             \and
             Department of Physics and Astronomy, University of Turku, FI-20014 Turku, Finland
            \and
            Hvar Observatory, Faculty of Geodesy, University of Zagreb, Croatia
            \and
            Lockheed Martin Solar and Astrophysics Laboratory, Palo Alto, CA, USA
            \and
            CIRES, University of Colorado Boulder, Boulder, CO, USA
            }

   \date{}

 
  \abstract
  {Late on 2013 August 19, STEREO-A, STEREO-B, MESSENGER, Mars Odyssey, and the L1 spacecraft, spanning a longitudinal range of 222$^{\circ}$ in the ecliptic plane, observed an energetic particle flux increase. The widespread solar energetic particle (SEP) event was associated with a coronal mass ejection (CME) that came from a region located near the far-side central meridian from Earth's perspective. The CME erupted in two stages, and was accompanied by a late M-class flare observed as a post-eruptive arcade, persisting low-frequency (interplanetary) type II and groups of shock-accelerated type III radio bursts, all of them making this SEP event unusual.}
   {There are two main objectives of this study, disentangling the reasons for the different intensity-time profiles observed by the spacecraft, especially at MESSENGER and STEREO-A locations, longitudinally separated by only 15$^{\circ}$, and unravelling the single solar source related with the widespread SEP event.} 
   {The analysis of in situ data, such as particle fluxes, anisotropies and timing, and plasma and magnetic field data, is compared with the remote-sensing observations. A spheroid model is applied for the CME-driven shock reconstruction and the ENLIL model is used to characterize the heliospheric conditions, including the evolution of the magnetic connectivity to the shock.
   }
   {The solar source associated with the widespread SEP event is the shock driven by the CME, as the flare observed as a  post-eruptive arcade is too late to explain the estimated particle onset. The different intensity-time profiles observed by STEREO-A, located at 0.97 au, and MESSENGER, at 0.33 au, can be interpreted as enhanced particle scattering beyond Mercury's orbit. The longitudinal extent of the shock does not explain by itself the wide spread of particles in the heliosphere. The particle increase observed at L1 may be attributed to cross-field diffusion transport, and this is also the case for STEREO-B, at least until the spacecraft is eventually magnetically connected to the shock when it reaches $\sim$0.6 au.}
   {}
   \maketitle
%
\section{Introduction}
\label{sec:Introduc}
Solar activity provides the origin and environment for solar energetic particle (SEP) events, which are significant increases in the fluxes of charged particles measured in situ, mainly protons and electrons, being ejected from the Sun \citep{Reames2013}. These SEP events are associated with coronal mass ejections (CMEs) and X-ray flares, and all three are the result of the release of the huge energy stored in the magnetic field structures formed at the Sun. The release, acceleration, and transport of particles, and the relation between CMEs, flares, and SEP events have been studied over several decades \citep{Forbush1946,Cane1986,Cane1988} and are still under investigation \citep[e.g.][and references therein]{Lario2013,Lario2014,Lario2017,Kihara2020}. 

The SEP events are often classified into two categories, impulsive and gradual \citep{Cane1986, Reames1999}, on account of their observed properties, such as timescales, spectra, composition and charge states, and the associated radio bursts. However, some observations \citep{KocharovTorsti2002,Kallenrode2003, Papaioannou2016} indicate that there might be no distinct separation between the two groups, and they suggest that a continuum of event properties is present, including hybrid or mixed SEP events. According to the two-class paradigm, impulsive SEP events are primarily accelerated by processes related with flares and jets, and gradual SEP events are the result of particle acceleration in the corona and the solar wind by CME-driven shock waves. If the CMEs and their interplanetary (IP) counterparts, hence IP CMEs (ICMEs), are fast enough, the corresponding coronal and IP shocks that they drive can accelerate SEPs to high energies \citep[e.g.][and references therein]{Desai2016}. Thus, gradual SEP events tend to be intense, energetic, and spatially and temporally extended, reflecting particles that have been accelerated over long timescales \citep{Reames2013}.

During some gradual events, SEPs originating from the CME-driven shock are detected over a very wide range of heliolongitudes. These widespread SEP events have been extensively researched \citep[e.g.][]{Reames1996,Wibberenz2006,Lario2006,Lario2013,Lario2016,Dresing2012,Dresing2014,Papaioannou2014,Richardson2014,Gomez-Herrero2015,Paassilta2018,Xie2019} thanks to constellations of spacecraft widely distributed throughout the heliosphere, such as Helios 1, Helios 2, Ulysses, the \textit{SOlar and Heliographic Observatory} \citep[SOHO,][]{Domingo1995SOHO}, the \textit{Solar TErrestrial RElations Observatory} \citep[STEREO,][]{Kaiser2008STEREO}, and other planetary missions. The ever-changing configuration of the STEREO spacecraft is advantageous for the study of the longitudinal variations of SEP events near 1 au, together with, for instance, SOHO located at L1. Several studies have clearly demonstrated that SEPs can arrive at Earth from solar events anywhere on the far side of the Sun, despite their great distance from the footpoint of the nominal spiral magnetic field line connected to Earth \citep[e.g.][]{Dodson1969,Torsti1999,Cliver2005, Dresing2012,Richardson2014,Gomez-Herrero2015}. 

The particular evolving connection of the observer to the CME-driven shock results in different SEP intensity profiles \citep[e.g.][]{Cane1988,Reames1999,CaneLario2006}. It is also known that a driven shock is often fastest and strongest near the `nose', while it is expected to show a decrease in shock speed and strength towards  the `flank' areas \citep{Kallenrode1993, Neugebauer2013}. The term nose of the shock is defined in opposition to the flanks, but it may lead to an over-simplistic depiction of the complexity, width, and size of the shock surface \citep[][]{Reames2013}. Therefore, the knowledge of the CME and CME-driven shock parameters, such as geometry and speed, plays a key role in understanding SEP events \citep[e.g.][]{Kallenrode1997,Lario2013,Desai2016, Kouloumvakos2019}. The angular extent of IP shocks can be investigated using multi-point in situ data, while the size of the coronal shocks can be indirectly deduced via remote-sensing observations. However, as the information provided by a single coronagraph image is an approximation of the real structure, the multi-view 3D reconstruction of the coronal shock geometry to correct projection effects is preferable  \citep[e.g.][]{Lario2014,Lario2016}. 

The extent of the coronal and IP shocks may sometimes explain the extremely wide longitudinal spread of particles \citep[e.g.][]{Lario2016}. However, in other SEP events, some additional physical processes allowing particle propagation perpendicularly to the average direction of the IP magnetic field (IMF) are required to explain the observations. These processes could include scattering by IP magnetic field turbulence \citep[e.g.][]{Dresing2012,Droege2016}, field line meandering  \citep[e.g.][and references  therein]{Laitinen2015b}, and deviations from the ideal IP magnetic field line spiral structure \citep[e.g.][]{Richardson1991,RichardsonCane1996,GomezHerrero2007,Chollet2010}. 

As an example of deviations from ideal IP magnetic field configuration, CMEs prior to an SEP event can distort the nominal transport conditions between the Sun and the different spacecraft \citep[e.g.][]{LarioKarelitz2014}. To have an insight into the possible solar wind structures present in the heliosphere in which CMEs and SEPs propagate, preceding eruptions generally have to be included in global 3D magnetohydrodynamic (MHD) simulations \citep[e.g.][]{Bain2016,Dumbovic2019}. This characterization of the heliosphere is   more accurate when the background solar wind is well replicated and if multi-point coronagraph observations are used to infer CME parameters \citep[e.g.][]{Lee2013,Mays2015}.

On 2013 August 19, an unusual large and widespread SEP event occurred that could be seen by different spacecraft located in the inner heliosphere, spanning a longitudinal range of 222$^{\circ}$ in the ecliptic plane. The SEP event was  observed by \textit{MErcury Surface Space ENvironment GEochemistry and Ranging} \citep[MESSENGER,][]{Solomon2007MESSENGER}, STEREO-A, and \textit{Mars Odyssey} \citep{Saunders2004MarsOdyssey},  all three being nominally well connected to the source, and also by STEREO-B and spacecraft at L1, such as SOHO and the \textit{Advanced Composition Explorer} \citep[ACE,][]{Stone1998ACE}, all having large longitudinal separation between the solar source and the footpoint of the respective nominal field lines. The SEP origin was associated with a wide and fast CME, erupting near the far-side central meridian from Earth's perspective.

The term `unusual' is linked to several circumstances. Firstly, the single CME, driving a single shock associated with the SEP event, erupted in two stages. Secondly, the observed radio dynamic spectrum associated with the SEP event presented groups of shock-accelerated (SA) type III bursts from $\sim$50 minutes before to $\sim$40 minutes after the release of particles. Thirdly, the flare observed as a post-eruptive arcade was not likely related with the particle acceleration. 

In addition to the unusual features, the difference in intensity profiles observed by STEREO-A (delayed and gradual) and MESSENGER (prompt and sharp), which were both relatively well connected to the source and only separated by 15$^{\circ}$ in heliolongitude, provides a good opportunity for the analysis of the SEP IP propagation between $\sim$0.3 au and $\sim$1 au.

To shed some light on which physical mechanisms are behind this unusual widespread SEP event, there are two specific objectives in this study: firstly, to identify the origin and evolution of the single solar source related with the widespread SEP event; and secondly, to disentangle the acceleration and propagation conditions that might be responsible for the SEP time profiles observed at different spacecraft, especially at the locations of MESSENGER and STEREO-A. 
 \begin{figure*}
\centering
   \includegraphics[width=12.46cm,height=14.91cm]{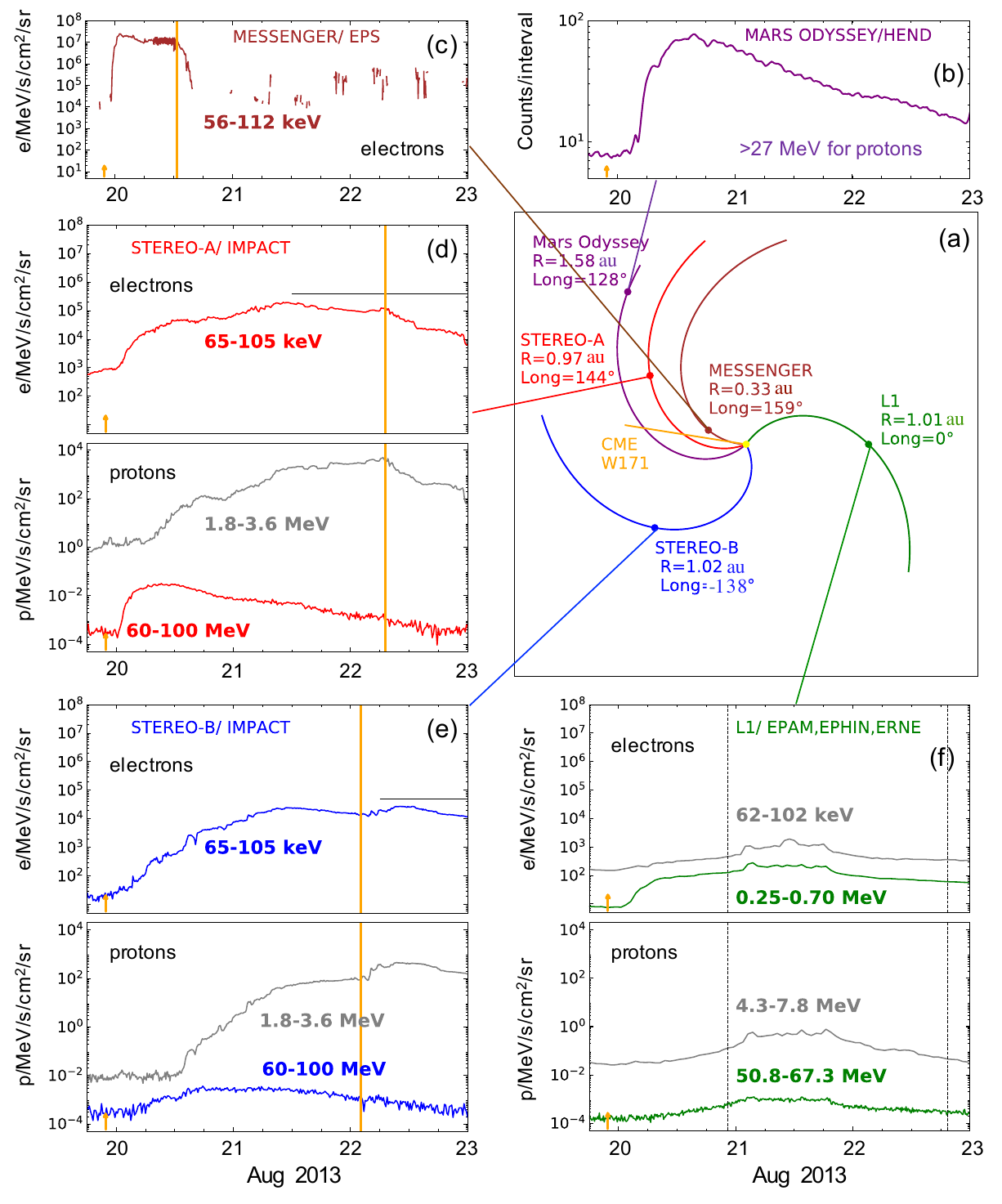}
     \caption{Overview of the widespread SEP event observed on 2013 August 19 at different spacecraft locations. (a) View from the north heliographic pole showing the locations of STEREO-A (red), near-Earth observers (L1; green), STEREO-B (blue), MESSENGER (brown) and Mars Odyssey (purple) on 2013 August 20. R and Long indicate the heliocentric distance and heliographic longitude of each observer, respectively, in the HEEQ coordinates. Also shown are the nominal IP magnetic field lines connecting each spacecraft with the Sun (yellow circle at the centre, not to scale), considering the mean solar wind measured around the solar release time of the SEP event (for MESSENGER and Mars Odyssey  the same speed as measured at STEREO-A was assumed). The orange line indicates the longitude of the parent CME apex direction. (b) Particle counting rate measured by Mars Odyssey in the higher channels of the inner scintillator with the background subtracted and smoothed with a third-order polynomial regression. (c) Near-relativistic electron intensities measured by MESSENGER. (d), (e), and (f) Near-relativistic electron (top) and proton intensities measured by STEREO-A, STEREO-B, and ACE and SOHO, respectively. The orange arrows on the horizontal axis indicate the occurrence time of the parent CME, as observed in  EUV  data, and the vertical orange lines show the passage of the IP shocks related with the SEP event. The shocks arriving the Earth that are not associated with the parent solar source are indicated with vertical grey dashed lines in (f). The horizontal lines in (d) and (e) represent a period with significant proton contamination of the electron fluxes observed by both STEREO/SEPT.}.
     \label{fig:SEP_SC_Position}
\end{figure*}
\begin{table*}[ht]
\begin{minipage}{\textwidth} 
\caption{Spacecraft locations, magnetic field line lengths and footpoint locations, and connection angles with respect to the AR}
\label{table:Magnetic footpoint location}
\begin{tabularx}{1\textwidth}{ccccccccccccccc} 
\hline
\hline
 \multicolumn{4}{c}{Spacecraft location}& Solar Wind & Field line &\multicolumn{4}{c}{Field line footpoint location}&Connection\\
\multicolumn{4}{c}{}& Speed &  Length&\multicolumn{4}{c}{}&Angle\\
 & \multicolumn{3}{c}{Location\textsuperscript{a}}& In situ/ENLIL\textsuperscript{b} & Parker/ENLIL\textsuperscript{d}& \multicolumn{2}{c}{Parker Spiral\textsuperscript{e}} & \multicolumn{2}{c}{ENLIL\textsuperscript{f}}&Parker/ENLIL\textsuperscript{g}\\
s/c&R&Lon&Lat& Vsw&L&Lon&Lat&Lon& Lat&$\Delta$$\phi$&\\
&(au)&(deg)&(deg)&(km/s)&(au)&(deg)&(deg)&(deg)&(deg)&(deg)\\
(1)&(2)&(3)&(4)&(5)&(6)&(7)&(8)&(9)&(10)&(11)\\
 \hline
STEREO-B & 1.02& -138& -7& 332/377 & 1.24/1.26&-64 & -7  & -87 & -6 &125/102 \\
MESSENGER & 0.33& 159& 1 & 418\textsuperscript{c}/301 & 0.31/0.31&177 & 1 & 174 & 1 &6/3\\
STEREO-A &0.97&144& -4 & 418/315 & 1.08/1.14& -160 & -4  & -164 & -5&29/25\\
MARS &1.58& 128& -1 & 418\textsuperscript{c}/322 & 2.09/2.26& -141 & -1  & -145 & -3&48/44\\
EARTH& 1.01& 0& 7 & 446/477 & 1.12/1.09&54 & 7  & 33& 6&-117/-138 \\

\hline
\\
\end{tabularx}
\footnotesize{ \textbf{Notes.}

\textsuperscript{a} STEREO location in HEEQ coordinates at 23:22 universal time (UT) on 2013 August 19. MARS and MESSENGER location in HEEQ coordinates at 23:00 UT on 2013 August 19.

\textsuperscript{b} One-hour averaged solar wind speed measured at estimated particle solar release time. ENLIL values are taken from simulation on 2013 August 19 at 23:00 UT.

\textsuperscript{c} Mercury and Mars solar wind speeds from STEREO-A.

\textsuperscript{d} Length calculated from 6 (5) R\textsubscript{$\odot$} for Parker spiral (ENLIL) magnetic field lines (as the probable heliospheric particle release height), assuming a radial extension from 21.5 R\textsubscript{$\odot$} to 5 R\textsubscript{$\odot$} in the case of ENLIL modelling.

\textsuperscript{e} Footpoint location at 6 R\textsubscript{$\odot$}.

\textsuperscript{f} Footpoint location radially extended from 21.5 R\textsubscript{$\odot$} to 5 R\textsubscript{$\odot$}.

\textsuperscript{g} The positive sign means westward location with respect to the CME apex direction (171$^{\circ}$).

}
\end{minipage}
\end{table*}
\begin{figure*}[htbp]
\centering
    \resizebox{\hsize}{!}{\includegraphics{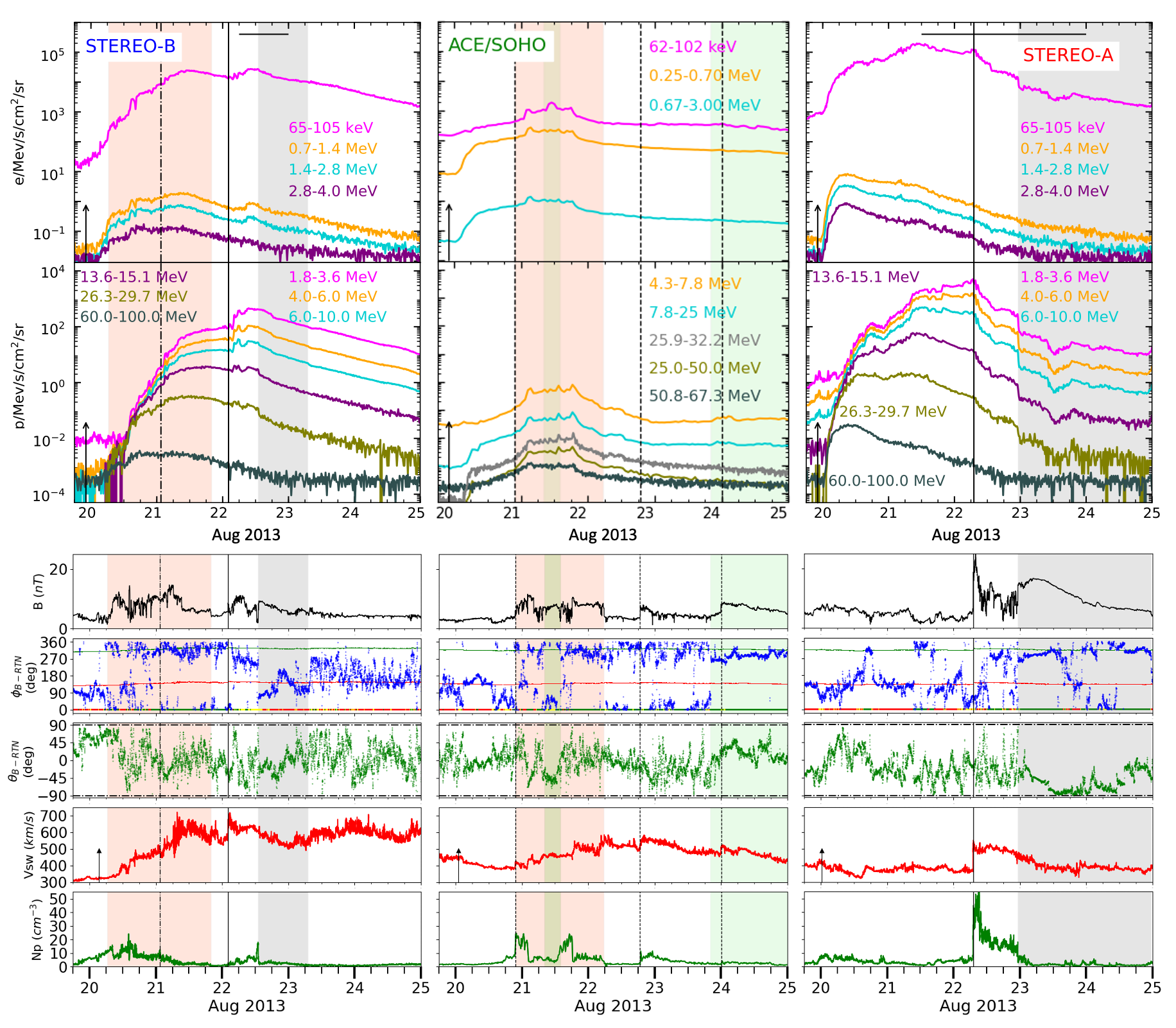}}
     \caption{In situ SEP time profiles and plasma and magnetic field observations by STEREO-B, ACE/SOHO, and STEREO-A. Top: Energetic electron and proton temporal profiles observed by, from left to right, STEREO-B, ACE/SOHO, and STEREO-A from comparable energy channels. The CME eruption time is represented by the arrow on the lower x-axis, while the horizontal lines on the upper x-axis of electron panels represent periods of proton contaminating the measured electron fluxes observed by both STEREO. A stream interface observed by STEREO-B is shown as a dash-dotted line, while the salmon shaded areas in STEREO-B and L1 indicate SIRs. The vertical solid lines and grey shaded areas indicate the IP shocks and ICME transits, respectively, associated with the SEP event, based on STEREO ICME list\textsuperscript{\ref{footnote Jian list}}. The IP shocks and ICMEs observed near the Earth, not related with the SEP event, are indicated with vertical dashed lines and green shaded areas, and are based on the SOHO CELIAS/MTOF Proton Monitor shock list\textsuperscript{\ref{footnote IP shock PM}} and the near-Earth ICME list\textsuperscript{\ref{footnote Richardson list}}, respectively. Bottom: In situ plasma and magnetic field observations by STEREO-B (left), ACE (center), and STEREO-A (right). The panels present, from top to bottom, the magnetic field magnitude, the magnetic field azimuthal and latitudinal angles, $\phi$\textsubscript{B-RTN} and $\theta$\textsubscript{B-RTN}, the solar wind speed, and the proton density, where RTN stands for radial-tangential-normal coordinates \cite[e.g.][]{Hapgood1992}. The coloured lines in the $\phi$\textsubscript{B-RTN} angle panel indicate the in situ magnetic field polarity, estimated from the magnetic field azimuth. Red and green colours denote inward and outward polarity, respectively. The lower band shows the observed magnetic field polarity, where the yellow intervals represent the periods with the magnetic field oriented close to perpendicular to the nominal Parker spiral. The respective relativistic electron onset times are marked with the arrow in the solar wind proton panel (IP structures as described in top panels).}
     \label{fig:proton_electron_fluxes_solar_wind}
\end{figure*}

Section \ref{sec:INSTRUM} introduces the instrumentation used in this paper. Section \ref{sec:SEPObservations} presents the SEP event, showing the energetic particle fluxes, anisotropies, spectra and timing at different locations, as well as the magnetic connectivity. Section \ref{sec:OBSER_remote-sensing} examines the SEP parent solar source, including remote-sensing observations, such as extreme ultraviolet (EUV), coronagraph, soft-X ray, and radio emission. It also presents the reconstruction of the CME, using the graduated cylindrical shell (GCS) model developed by \citet[]{Thernisien2006GCS} and \citet{Thernisien2011}, and the associated CME-driven shock, applying an spheroid model based on \cite{Olmedo2013}. Section \ref{sec:heliospheric_conditions} presents the heliospheric conditions that might affect the particle transport. We analyse the in situ solar wind along with the Wang-Sheeley-Arge (WSA)-ENLIL + Cone model \citep[hereafter ENLIL model;][]{Odstrcil1996,Arge2000ENLIL,Odstrcil2003,Arge2004ENLIL} simulation. Section \ref{sec:discussion} summarizes the SEP event and presents the findings of the research. Finally, Sect. \ref{sec:Conclusions} outlines the main conclusions. 

\section{Instrumentation} 
\label{sec:INSTRUM}
The comprehensive study of the SEP event requires the analysis of observations from a wide range of instrumentation on board different spacecraft. We used data from STEREO-A, STEREO-B, MESSENGER, ACE, SOHO, Mars Odyssey, Wind \citep{Szabo2015}, the \textit{Solar Dynamics Observatory} \citep[SDO,][]{Pesnell2012}, the \textit{Geostationary Operational Environmental Satellites} \citep[GOES,][]{Garcia1994}, and also from ground-based experiments, such as the \textit{Learmonth spectrograph}\footnote{\url{http://www.sws.bom.gov.au/Solar/3/2}} and the \textit{Bruny Island Radio Spectrometer} \citep[BIRS,][]{Erickson1997}. Remote-sensing observations of the CME and the activity on the solar surface were provided by the Atmospheric Imaging Assembly \citep[AIA,][]{Lemen2012} on board SDO, the C2 and C3 coronagraphs of the Large Angle and Spectrometric COronagraph \citep[LASCO,][]{Brueckner1995} instrument on board SOHO, and the Sun Earth Connection Coronal and Heliospheric Investigation \citep[SECCHI,][]{Howard2008SECCHI} instrument suite on board STEREO. In particular, the COR1 and COR2 coronagraphs and the Extreme Ultraviolet Imager \citep[EUVI,][]{Wuelser2004}, part of SECCHI suite, were utilized. Radio observations were provided by the S/WAVES \citep{Bougeret2008_S/WAVES} investigation on board STEREO, the WAVES \citep{Bougeret1995} experiment on board Wind, and the Learmonth and the BIRS spectrometers. The X-Ray telescopes of the GOES satellites were also used.
In situ energetic particle observations were provided by the Solar Electron and Proton Telescope \citep[SEPT,][]{Mueller2008SEPT}, the Low-Energy Telescope  \citep[LET,][]{Mewaldt2008LET}, and the High-Energy Telescope \citep[HET,][]{vonRosenvinge2008HET} on board STEREO \citep[all of them part of the IMPACT instrument suite,][]{Luhmann2008IMPACT}; the Electron Proton and Alpha Monitor \citep[EPAM,][]{Gold1998EPAM} on board ACE; the Electron Proton Helium INstrument (EPHIN), part of the Comprehensive Suprathermal and Energetic Particle Analyzer \citep[COSTEP,][]{Mueller-Mellin1995COSTEP} and the Energetic Relativistic Nuclei and Electron Instrument \citep[ERNE,][]{Torsti1995ERNE} on board SOHO; the 3D Plasma and Energetic Particle Investigation (3DP)\footnote{\url{http://sprg.ssl.berkeley.edu/wind3dp/}} on board Wind; the Energetic Particle and Plasma Spectrometer \citep[EPPS,][]{Andrews2007EPPS} on board MESSENGER; and the High Energy Neutron Detector (HEND), part of the Gamma Ray Spectrometer Suite \citep[GRS,][]{Boynton2004GRS}, on board Mars Odyssey, which is sensitive to charged particles although it was not primarily designed to measure them \citep[e.g.][]{Zeitlin2010}. Solar wind plasma and magnetic field observations were provided by the Plasma and Suprathermal Ion Composition \citep[PLASTIC,][]{Galvin2008} investigation and the Magnetic Field Experiment \citep{Acuna2008} on board STEREO; the Magnetometer Instrument \citep{Anderson2007} on board MESSENGER; and the Magnetic Field Experiment \citep{Smith1998ACEMag}, the Solar Wind Electron Proton Alpha Monitor \citep[SWEPAM,][]{McComas1998SWEPAM} and the Solar Wind Ion Composition Spectrometer \citep[SWICS,][]{Gloeckler1998} on board ACE. The CDAW SOHO LASCO CME catalogue\footnote{\url{https://cdaw.gsfc.nasa.gov/CME_list/}} \citep{Yashiro2004} provided the CME timing from Earth's perspective. Interplanetary structures were classified using the STEREO level 3 event lists\footnote{\url{https://stereo-ssc.nascom.nasa.gov/data/ins_data/impact/level3/}\label{footnote Jian list}} \citep{Jian2018,Jian2019} maintained by L. Jian, the near-Earth ICME list provided by I. Richardson and H. Cane\footnote{\url{http://www.srl.caltech.edu/ACE/ASC/DATA/level3/icmetable2.htm}\label{footnote Richardson list}} \citep{Richardson2010}, the SOHO CELIAS/MTOF Proton Monitor shock list\footnote{\url{http://umtof.umd.edu/pm/FIGS.HTML}\label{footnote IP shock PM}}, the IP shocks catalogue maintained by the University of Helsinki\footnote{\url{http://www.ipshocks.fi/}}, and the ICME catalogue at Mercury from the University of New Hampshire\footnote{\url{http://c-swepa.sr.unh.edu/icmecatalogatmercury.html}}. The magnetograms from Global Oscillations Network Group \citep[GONG,][]{Harvey1996GONG} are available from the National Solar Observatory website\footnote{\url{https://gong.nso.edu/data/magmap/index.html}}. 

 \begin{figure*}
\centering
    \resizebox{\hsize}{!}{\includegraphics{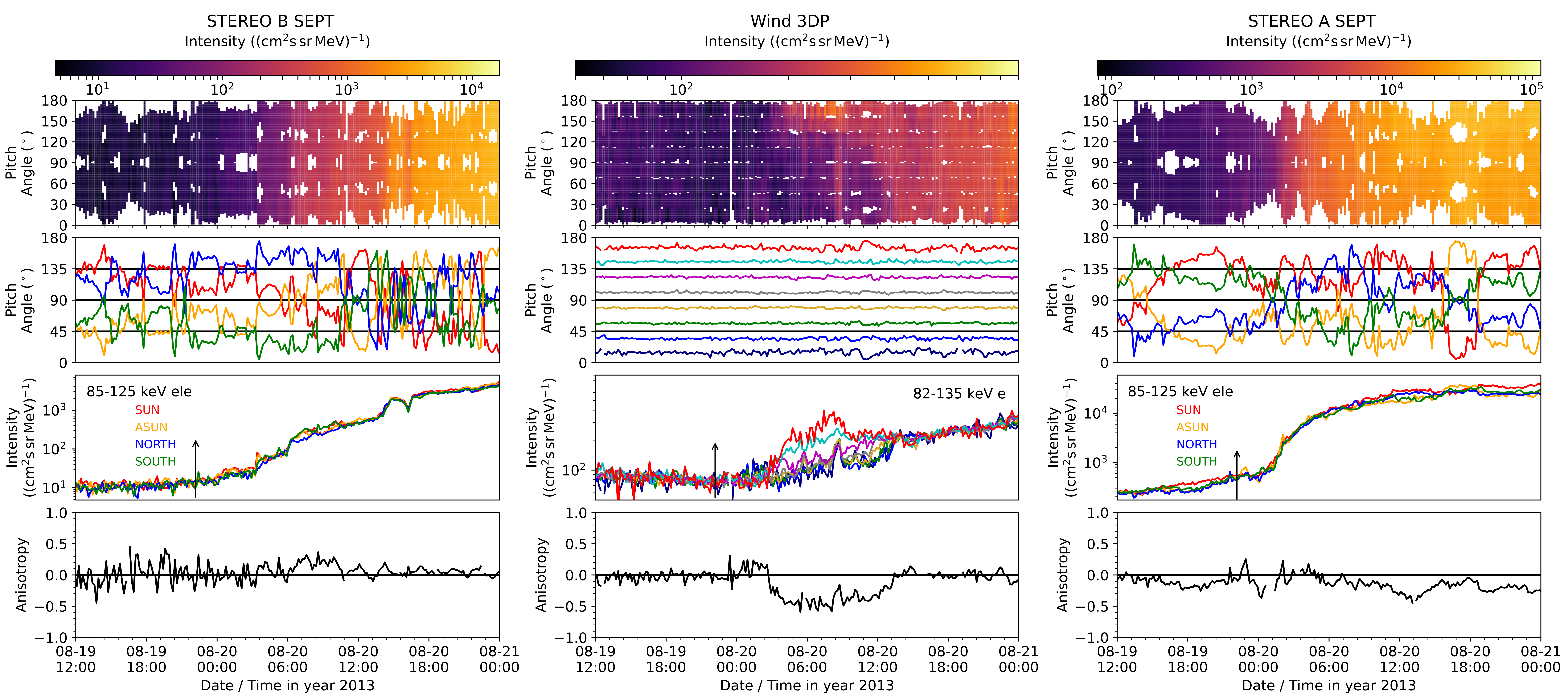}}
     \caption{Anisotropy and intensity time profiles of near-relativistic electrons as observed by, from left to right, STEREO-B/SEPT, Wind/3DP, and STEREO-A/SEPT. The first row of panels shows pitch-angle dependent intensity distributions, according to the colour intensity code shown on the upper horizontal bar. The second row of panels indicates the pitch angles covered by the centre of the four telescopes of STEREO/SEPT (sun in red, anti-sun in orange, north in blue, and south in green), and each of the eight pitch-angle bins of the Wind/3DP. The third row of panels indicates the electron flux intensities measured by each of the telescopes or bins, where the vertical arrows indicate the CME onset time. The fourth row of panels represents the first-order anisotropy values, in the range [-3, 3] \citep[e.g.][]{Dresing2014}, but zoomed in on the  [-1, 1] interval. The energies examined are 85-125 keV for STEREO and 82-135 keV for Wind.}
     \label{fig:anisotropies_electrons}
\end{figure*}
\section{SEP event: In situ observations and analysis}
\label{sec:SEPObservations}

 By the end of 2013 August 19 and beginning of August 20, STEREO-A, STEREO-B, and L1 saw a moderate and gradual increase in both near-relativistic electrons, from 65 to 105 keV, and in protons up to 50 MeV energies, while MESSENGER detected a sharp and prompt increase in near-relativistic electrons up to the 708 keV channel, and so did Mars-Odyssey, at least in protons energies above 27 MeV; the details about the corresponding particle energies can be found in \cite{Zeitlin2010}. The configuration of spacecraft locations in the heliocentric Earth equatorial (HEEQ) coordinates \citep[e.g.][]{Thompson2006} early on 2013 August 20, and the in situ measured fluxes of energetic particles from late on 2013 August 19 to August 22 at the different spacecraft locations are shown in Fig. \ref{fig:SEP_SC_Position}. Figure \ref{fig:SEP_SC_Position}(a) shows that STEREO-A, STEREO-B, and L1 are located at heliocentric distances near 1 au, while MESSENGER is situated at 0.33 au, and Mars Odyssey is located at 1.58 au. The parent CME apex direction (orange line in Fig.\ref{fig:SEP_SC_Position}(a)) is located W171$^{\circ}$ as seen from Earth, W12$^{\circ}$ as seen from MESSENGER, W27$^{\circ}$ as seen from STEREO-A, W43$^{\circ}$ as seen from Mars Odyssey, and E51$^{\circ}$ as seen from STEREO-B. The nominal Parker spiral magnetic field lines connecting the spacecraft to the Sun show that MESSENGER, STEREO-A, and Mars Odyssey (separated by 31$^{\circ}$) are relatively well connected to the source; and we expected to see similar flux increase behaviours, but this did not occur (Fig. \ref{fig:SEP_SC_Position}(b) to (d)).
 
 To characterize the SEP event at different radial and longitudinal locations, in this section we analyse further the solar energetic particles fluxes and anisotropies at 1 au (determining a reliable first-order anisotropy for MESSENGER at 0.33 au is not possible, due to the lack of sunward-pointing field of view) and the particle timing and magnetic connectivity for different spacecraft. We also examine the particle spectra at STEREO. 

\begin{figure*}
\centering
    \resizebox{\hsize}{!}{\includegraphics{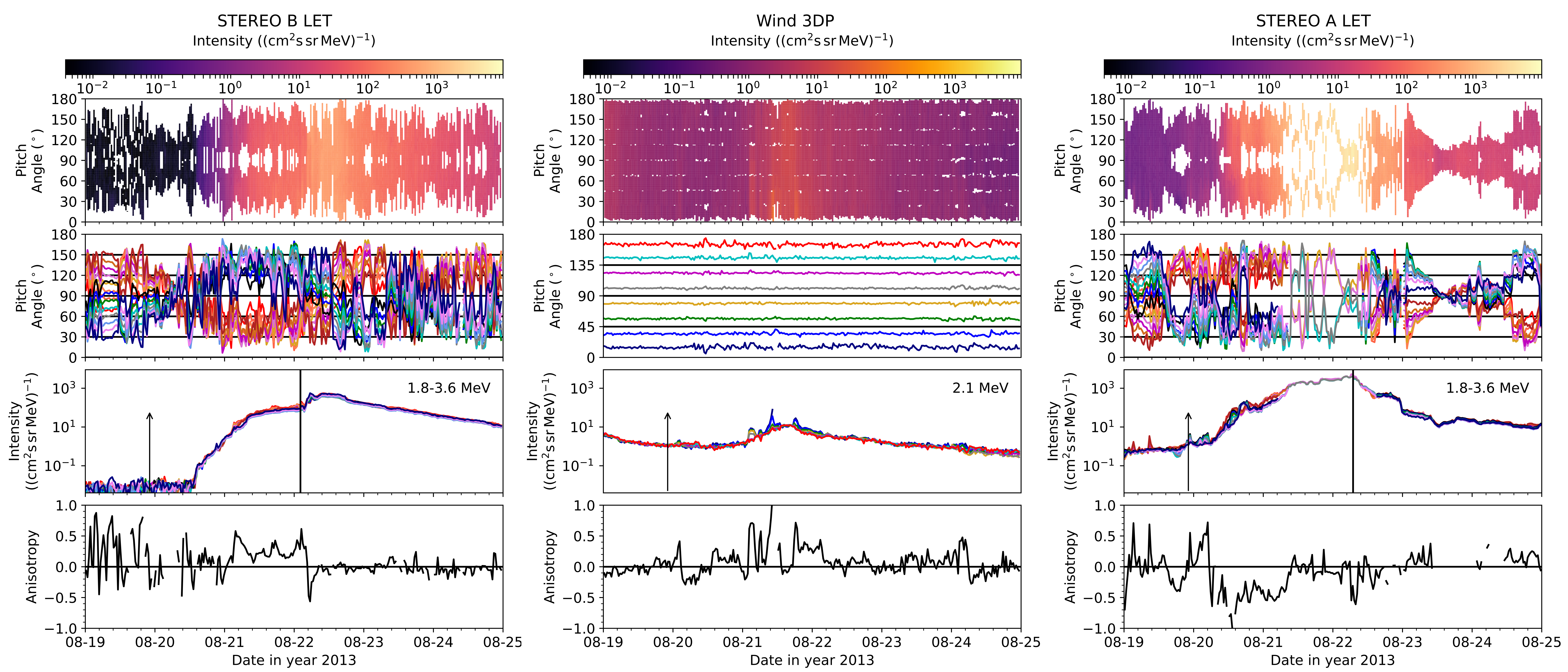}}
     \caption{Anisotropy and intensity time profiles of protons as observed by, from left to right, STEREO-B/LET, Wind/3DP, and STEREO-A/LET. The panels show almost the same information as in Fig. \ref{fig:anisotropies_electrons}, but with 16 sectors in STEREO/LET,  eight front-side (reddish colours)  and eight back-side sectors (bluish colours). The energies examined are 1.8-3.6 MeV for both STEREO and 2.1 MeV for Wind. The vertical lines indicate the CME-driven shock arrival to the spacecraft.}
     \label{fig:anisotropies_protons}
\end{figure*}

\subsection{Magnetic connectivity}
\label{sec:magneticFootpoint}
A fundamental parameter for interpreting the SEP event profiles at different locations is the longitudinal separation between the solar source and the footpoint of the IMF lines connecting to the respective observer. The location and magnetic connectivity around the estimated SEP onset time of the different spacecraft is shown in Fig. \ref{fig:SEP_SC_Position}(a) and detailed in Table \ref{table:Magnetic footpoint location}. Columns (2) to (4), present the spacecraft locations\footnote{\url{https://stereo-ssc.nascom.nasa.gov/cgi-bin/make_where_gif}} \footnote{\url{https://omniweb.gsfc.nasa.gov/coho/helios/heli.html}} and, Col. (5) shows the solar wind speed derived from in situ measurements. The nominal magnetic field line length, calculated from $\sim$6 R\textsubscript{$\odot$} (as the probable particle release height discussed in Sect. \ref{sec:Analysis_VDA}) to the respective spacecraft locations, is shown in Col. (6). Columns (7) and (8) present the footpoint location at 6 R\textsubscript{$\odot$}, given by the nominal Parker magnetic field lines connecting the respective spacecraft. The connection angle ($\Delta\phi$), which is  the longitudinal separation between the CME apex direction and the footpoint of the magnetic field line connecting to the corresponding spacecraft, based on the Parker spiral, is shown in Col. (11). MESSENGER, with $\Delta\phi$= +6{$^{\circ}$} and STEREO-A, with $\Delta\phi$= +29{$^{\circ}$}, are both nominally well connected to the source; the positive sign means westward location with respect to the CME apex direction. This is not the case for STEREO-B and the Earth, which have $\Delta\phi$ over 100{$^{\circ}$}.

The synoptic map from GONG on August 19 at 23:14 universal time (UT)\footnote{\url{ftp://gong2.nso.edu/oQR/zq5/201308/mrzq5130819/mrzq5130819t2314c2140_097.gif}} (not shown) indicates that the STEREO-A and MESSENGER footpoints are situated close to each other and close to the CME eruption location in heliolongitude, but STEREO-A might be separated from the parent active region (AR) by two Heliospheric Current Sheet (HCS) crossings. 

 \subsection{Energetic particle fluxes at 1 au}
\label{sec:OBSER_SEP_Fluxes}
The upper part of Fig.  \ref{fig:proton_electron_fluxes_solar_wind} shows an overview of the energetic particle fluxes observed by STEREO-A (right panels), STEREO-B (left panels), and ACE/SOHO (central panels), from late on 2013 August 19 to August 24. The upper and lower panels display the flux of electrons and protons at different energies: one-hour averaged for EPAM and EPHIN, ten-minute averaged for ERNE, SEPT-sun telescope and LET-summed\footnote{\url{http://www.srl.caltech.edu/STEREO/docs/LET_Level1.html}}, and fifteen-minute averaged for HET. All three spacecraft show a gradual increase in electrons and protons, but the particle intensities observed by the L1 spacecraft show an even more gradual increase than STEREO-B. STEREO-A presents higher intensities than STEREO-B for all comparable energies and species, while SOHO shows intensities similar to those of STEREO-B for electrons in the range of 0.7-3.0 MeV, but lower intensities for protons. However, if cross-calibration is taken into account (Fig. 2 in \citealt{Lario2013} and Fig. 3 in \citealt{Richardson2014}), STEREO-B registers higher intensities than SOHO for all comparable energies and species. STEREO electron fluxes below 500 keV (pink line) are susceptible to proton contamination during periods of enhanced sub-MeV proton flux, for example the periods indicated with a horizontal line in both STEREO. From the middle of August 21 to the end of August 23, protons of 500 keV reach the STEREO-A spacecraft and contaminate the electron channels. However, the first part of the event, where the main increase is seen, is clean because electrons arrive much more quickly than the slower low-energy protons causing the contamination. At STEREO-B, the time with proton contamination is observed after the arrival of an IP shock, shown as a vertical solid line. 

\subsection{Energetic particle anisotropies at 1 au}
\label{sec:Anisotr}

The directional information provided by different sectors in STEREO/SEPT and LET, and in Wind/3DP, can be used to obtain the electron and proton pitch-angle distribution, where the pitch angles are defined relative to the locally measured magnetic field direction. We   analysed the pitch-angle information available from STEREO and Wind, from 2013 August 19 to August 25. 

Figure \ref{fig:anisotropies_electrons} shows electron anisotropies observed by STEREO-A (right panel), Wind (middle panel), and STEREO-B (left panel) around the SEP onset time (from midday of 2013 August 19 to the end of August 20) at energies of 85 to 125 keV  for STEREO, and 82 to 135 keV for Wind. We note that the fourth row of panels represents the first-order anisotropy values, in the range [-3, 3] \citep[e.g.][]{Dresing2014}, but zoomed in on the [-1, 1] interval. High positive (negative) values mean that the pitch-angle distribution reaches a clear maximum in intensity at pitch angle 0{$^{\circ}$} (180{$^{\circ}$}), where an anisotropy value close to 0 means isotropic behaviour. In spite of its relatively small connection angle ($\Delta\phi$= +29{$^{\circ}$}), STEREO-A (right panels) observes limited anisotropy during the early phase of the event (although the pitch-angle coverage is not optimal), when the largest anisotropies are usually observed in well-connected SEP events \citep[e.g.][]{He2011,Dresing2014}. The small anisotropy observed after the CME eruption, indicated by the arrow, is due to an increase in electrons at the pitch-angle sector closer to 180{$^{\circ}$} (red line in the second panel), which corresponds to particles propagating anti-sunwards, given the negative magnetic polarity. STEREO-B (left panels) does not show clear anisotropies around the onset of the particles either, which is not surprising given its large connection angle ($\Delta\phi$= +125{$^{\circ}$}). However, after the onset phase, at the time of the stream interaction region (SIR) observed by the spacecraft (corresponding to the salmon shaded area in Fig. \ref{fig:proton_electron_fluxes_solar_wind}), STEREO-B observes a slight increase in electrons near pitch angle 0{$^{\circ}$}. This increase corresponds to particles propagating anti-sunwards due to the positive magnetic field polarity. The middle panels of Fig. \ref{fig:anisotropies_electrons} show Wind/3DP electron anisotropies, which present a better pitch-angle coverage with eight sectors. The fourth panel shows a period with electron anisotropy that starts around $\sim$03:15 UT on August 20, around five hours later than the CME eruption associated with the main SEP event that originated on the far side of the Sun. This timing agrees with the brightening seen at $\sim$03:20 UT, on the west limb from Earth's perspective (AR3 in Table \ref{table:Active_Regions}), erupting well after the L1 relativistic 0.25-0.70 MeV electron onset ($\sim$01:00 UT). Thus, by reason of timing and magnetic connectivity, this anisotropic period is likely caused by this well-connected small event that overlaps with the main SEP event, and injects electrons in Wind spacecraft. However, during the particle onset and later in the rising phase (i.e. August 20 to August 23) Wind 82-135 keV electrons are contaminated by ions, as observed when compared  with the 553 keV proton intensities (not shown), so the data are not reliable. 

Figure \ref{fig:anisotropies_protons} shows two-minute averaged 1.8 to 3.6 MeV proton anisotropies for STEREO-A (right) and STEREO-B (left), and 2.1 MeV proton data for Wind (middle), from 2013 August 19 to August 25. The fourth row indicates that both STEREO-A (right panel) and STEREO-B (left panel) do not see strong anisotropies. However, from August 21 at 07:00 UT to August 22 at 12:00 UT there is a data interval at STEREO-A with reduced pitch-angle coverage that leads to a more uncertain anisotropy analysis. At the beginning of the event, after the CME onset indicated by the arrow in the third panel, STEREO-A shows a small proton anisotropy that it is not present in STEREO-B. STEREO-A presents a negative anisotropy most of the time (e.g. the period from mid August 20 to mid August 21),  meaning that the particles arrive at pitch angle 180{$^{\circ}$}, which is mainly covered by the front-side sectors (reddish colour), corresponding to particles propagating anti-sunwards. However, during a short period around August 20 17:00 UT, this pitch angle is covered by the bluish colours, which corresponds to particles arriving from the back. This behaviour can be due, for example, to an s-shaped magnetic field in the solar wind. STEREO-B observes a weak anisotropy period starting early on August 21. Later on it shows a sudden increase in the intensity profile associated with a shock arrival on August 22 at 02:09 UT. Around this time we can see how the anisotropy shifts from a positive to negative value due to a change in magnetic field polarity, as the particles are still arriving at pitch angle 0{$^{\circ}$}. The particle arrival direction changes from front-side sectors to rear-side sectors, which suggests that particles observed following the shock are arriving. STEREO-A has a more pronounced intensity increase associated with the shock arrival (third panel on the right) than STEREO-B. When the CME-driven shock arriving to STEREO-A and STEREO-B is the same, this behaviour could be explained by different parts of the shock having different acceleration efficiencies. Wind proton anisotropies (middle panels) are very low during the SEP onset time, due to the very large longitudinal separation between the CME eruption and the field line footpoint of the L1 spacecraft ($\Delta\phi$= -117{$^{\circ}$}). On August 21 there are three main spikes present at pitch angle 0{$^{\circ}$} that might not be associated with the main SEP event. As shown in Fig. \ref{fig:proton_electron_fluxes_solar_wind}, an ICME and corresponding IP shock, and an SIR arrive at the Earth that day. These anisotropies might be related with in situ particle acceleration or might be SIR-associated. 
\begin{figure}[htbp] 
   \resizebox{\hsize}{!}{\includegraphics{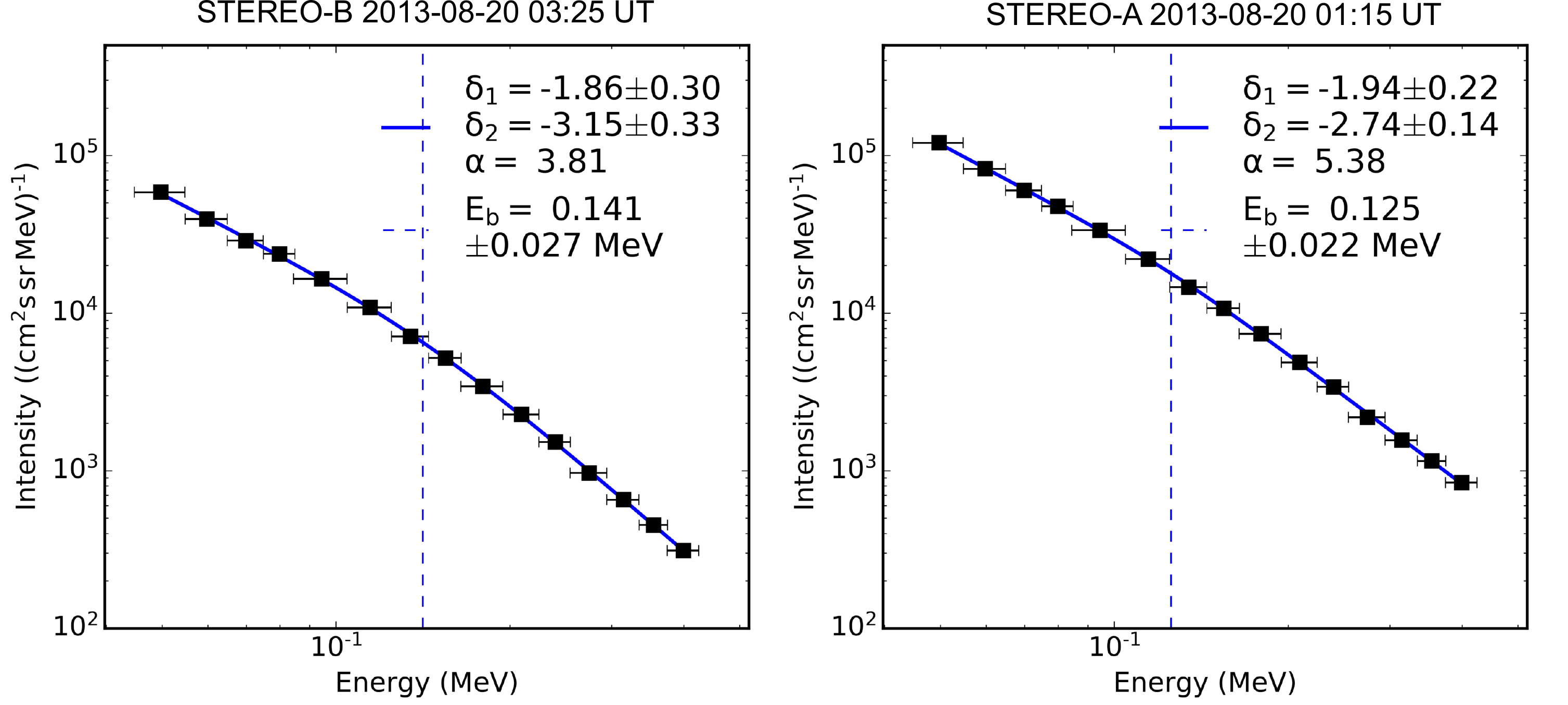}}
  \caption{Electron peak intensity spectra observed by STEREO-B (left) and STEREO-A (right). The legend shows the fit values: the spectral index below ($\delta_1$) and above ($\delta_2$) the spectral transition; $E_b$ (vertical dashed line); and $\alpha$, which determines the `sharpness' of the break \citep[][]{Strauss2020}.}
    \label{fig:spectra}
\end{figure} 
\begin{table*}[ht]
\caption{Energetic particle timing results and shock height at the particle release times}
\label{table_vda}
\begin{tabularx}{1\textwidth}{ccccccc} 
\hline
\hline
Spacecraft/Instrument/&  Effective path length& Estimated & \multicolumn{2}{c}{Shock height} & \multicolumn{2}{c}{$\theta$\textsubscript{Bn} at the}\\
 Particle Species& or IMF length\textsuperscript{a}&solar release time& \multicolumn{2}{c}{from Sun centre} & \multicolumn{2}{c}{cobpoint} 
 \\
&(au)&(UT)\textsuperscript{b}&\multicolumn{2}{c}{(R\textsubscript{$\odot$})\textsuperscript{c}}&\multicolumn{2}{c}{ (deg)\textsuperscript{d}}\\
&&&Parker& ENLIL&Parker&ENLIL\\

(1)&(2)&(3)&(4)& (5)&(6)&(7)\\
 \hline
STA/SEPT/HET\textsuperscript{e}  &3.55 $\pm$ 0.50 (VDA)  &23:28 $\pm$ 15 min    & 6.0 $\pm$ 2.0&5.7 $\pm$ 2.0 & 22 $\pm$ 2&26 $\pm$ 3\\
 \hline
STA/SEPT/ 65-105 keV e &1.14 $\pm$ 0.03 (TSA) &00:15 $\pm$ 10 min  & 9.5 $\pm$ 1.0&10.2 $\pm$ 1.0 & 19 $\pm$ 2&20 $\pm$ 2 \\
 \hline
STB/HET/0.7-1.4 MeV e  &1.26 $\pm$ 0.03 (TSA) &02:57 $\pm$ 15 min  & (27.5 $\pm$ 2.0)\textsuperscript{f} & -&-& - \\
 \hline
MESS/EPPS/71-112 keV e &0.31 $\pm$ 0.01 (TSA) &23:22 $\pm$ 5 min\textsuperscript{g}  & 5.9 $\pm$ 0.6&5.1 $\pm$ 0.6 & 5 $\pm$ 1&8 $\pm$ 1  \\
 \hline
SOHO/EPHIN/0.25-0.70 MeV e &1.09 $\pm$ 0.03 (TSA) &00:58 $\pm$ 60 min  & (15.6 $\pm$ 6.0)\textsuperscript{f} & -&-&- \\
 \hline
 \\
\end{tabularx}

\footnotesize{ \textbf{Notes.}

\textsuperscript{a} Magnetic field lines length for TSA analysis from ENLIL modelling. 

\textsuperscript{b} Solar release times late on 2013 August 19 and early on August 20, shifted by light propagation time to 1 au in order to compare with electromagnetic observations.

\textsuperscript{c} CME-driven shock heliospheric height at the cobpoint \citep[][]{Sanahuja1995} for STEREO-A and MESSENGER (calculated for Parker spiral and ENLIL), and at the shock nose for STEREO-B and SOHO. The range of heights includes the uncertainty in the estimated SEP release time, as discussed in \cite{Lario2016}. 

\textsuperscript{d} $\theta$\textsubscript{Bn}: The angle between the shock normal vector and the magnetic field lines at the cobpoint (estimated using Parker spiral and ENLIL, which are radially extended from 21.5 R\textsubscript{$\odot$}). The given uncertainties are rounded to one degree. They are based on the shock height deviation due to the uncertainty in the particle release time (the fluctuations in the magnetic field lines, due to waves and turbulence, or in the shock front geometry are not considered).

\textsuperscript{e} Energies used in VDA analysis: 0.065-2.8 MeV e \& 13.6-60 MeV p.

\textsuperscript{f} STEREO-B is not connected to the shock until later in time, and SOHO is never connected to the shock.

\textsuperscript{g} MESSENGER was pointing in the anti-Sun direction, thus the solar release time should be interpreted as an upper limit. 
}
\end{table*} 
\subsection{Electron spectra at STEREO}
\label{sec:spectra}

Following the method described by \cite{Dresing2020} and \cite{Strauss2020}, we determined the electron peak intensity spectra, as observed by STEREO/SEPT. Figure \ref{fig:spectra} presents the fitting, which shows double power-law shapes. The spectral index below and above the spectral transition is respectively $\delta_1=-1.94\pm0.22$  and $\delta_2=-2.74\pm0.14$ for STEREO-A, and $\delta_1=-1.86\pm0.30$  and  $\delta_2=-3.15\pm0.33$ for STEREO-B, where the spectral break or transition energy is $E_b=125 \pm 22$ keV for STEREO-A and $E_b=141 \pm 27$ keV for STEREO-B. In comparison   to the average values $\delta_1=-2.53$ and $\delta_2=-3.93$ of the large statistical sample studied by \cite{Dresing2020}, who investigated the electron events observed by STEREO from 2007 to 2018, the spectral indices observed in this event are clearly harder at both STEREO spacecraft. According to \cite{Strauss2020}, the observed spectral transitions are likely caused by pitch-angle scattering during transport through the interplanetary medium.

\begin{figure}[htbp] 
   \resizebox{\hsize}{!}{\includegraphics{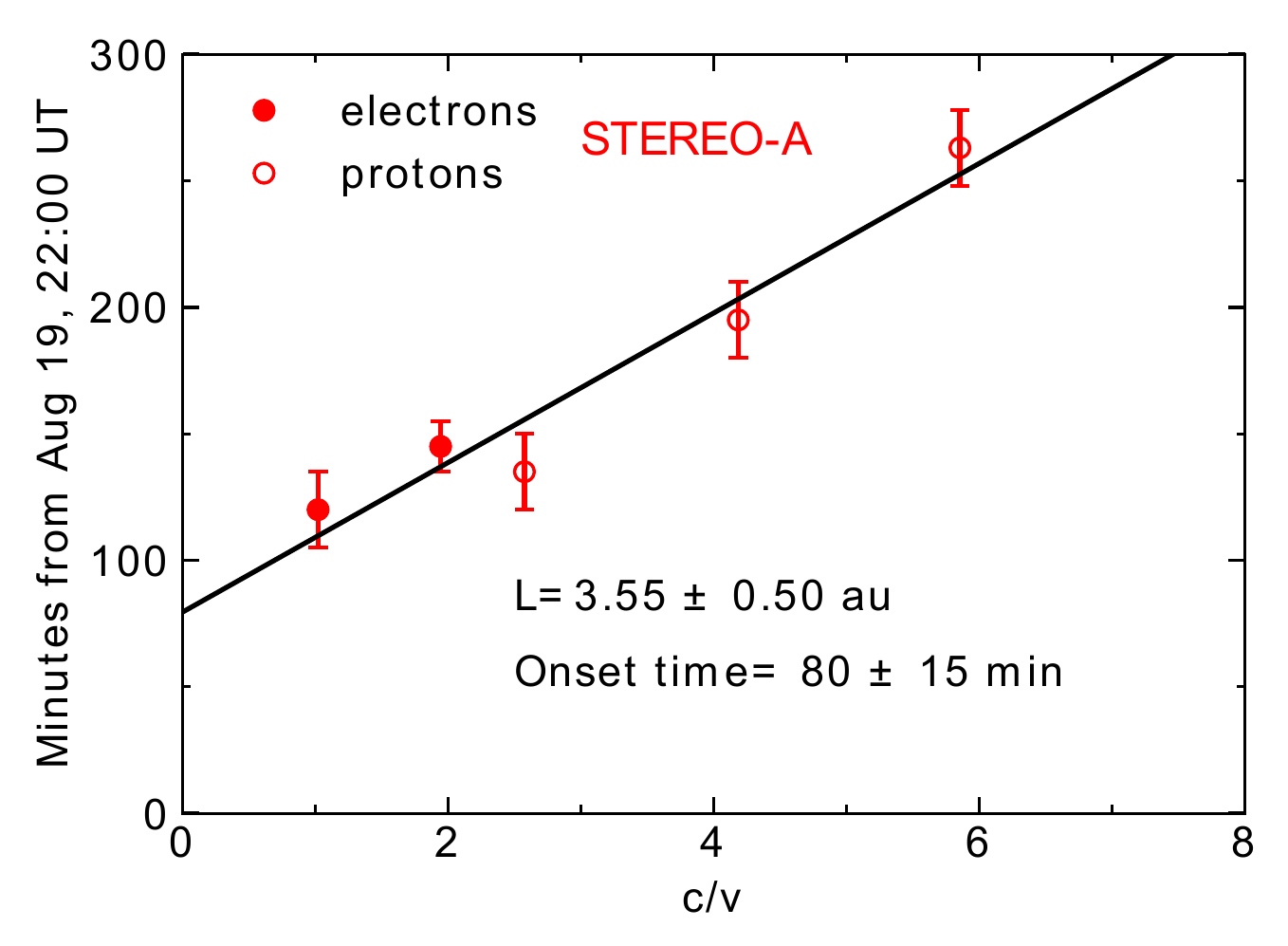}}
  \caption{Velocity dispersion analysis of the onset of the SEP event at STEREO-A. The horizontal and vertical axes correspond to the reciprocal of the particle velocities and onset times, respectively. The red points and circles respectively identify the electron and proton onsets at the corresponding velocities (energies), with the respective errors indicated. The black line is the linear regression fit to all points. The legend gives the effective path length (L) and the estimated release time (onset time) discussed in the text. }
    \label{fig.VDA_STEREOA}
\end{figure} 
\begin{table*}[htbp]
\caption{Active regions present in the period of study and signatures observed by the different instruments}
\label{table:Active_Regions}
\begin{tabular}{ccccccccc}
\hline
\hline
 AR & Location\textsuperscript{*} & Type of & Time \textsuperscript{***}& GOES & STEREO-A\textsuperscript{\#}& STEREO-B\textsuperscript{\#}& Wind\textsuperscript{\#}& BIRS\textsuperscript{\#}\\
 & &Activity\textsuperscript{**} &&XRS&WAVES&WAVES&WAVES\\ 

 \hline
 (1)&(2)&(3)&(4)&(5)&(6)&(7)&(8)&(9)\\
 \hline
AR1 & E53S18 & Brightening & 20/08 00:42&B7.5&-&-&-&- \\
 & (11827)& & &  &\\
AR2 & W66S04& Brightening& 19/08 21:58& C1.0 &-&-&-&- \\
 & (11818) &Jet & 20/08 00:10&B8.5&Type III (b\textsuperscript{+})&Type III (b\textsuperscript{+}) (oc.)& Type III (b\textsuperscript{+}) & Type III \\
 &  &   Brightening & 20/08 01:40 &-&Type III (c\textsuperscript{+})&-&Type III (c\textsuperscript{+})&-\\
  &  &  Brightening & 20/08 02:44&-&Type III (d\textsuperscript{+}) &Type III (d\textsuperscript{+}) (oc.)&Type III (d\textsuperscript{+}) & Type III\\
AR3 & W67S11 & Brightening & 20/08 03:20&B6.0&Type III (e\textsuperscript{+})&-&Type III (e\textsuperscript{+})&- \\
 & (11817/9) &  & &&&& \\
AR4 & W79S16 & Brightening & 20/08 01:25&C1.8&-&-&-&- \\
 & (11819) &  & && \\
AR5 & W91S19 & Jet & 19/08 22:55&B7.0&Type III (a\textsuperscript{+}) &Type III (a\textsuperscript{+}) (oc.) & Type III (a\textsuperscript{+})& Type III \\
 & (11814) &  &  && \\
AR6\textsuperscript{§} & W171N08 &1\textsuperscript{st} stage  & 19/08 21:20&-&- &-  &-&- \\
 & (11809) & 2\textsuperscript{nd} stage & 19/08 22:10 &-&-&- &-&- \\
&&&19/08 22:28 &-& Type II& Type II&-&-\\
 &  &  & & & SA type III (1->3\textsuperscript{+})& SA type III (1->3\textsuperscript{+})&-&- \\
 &  &  M flare starts  & 19/08 23:30&-&-&-&-&- \\
\hline
\\
\end{tabular}

\footnotesize{ \textbf{Notes.}

(*) Stonyhurst reference frame (AR number from National Oceanic and Atmospheric Administration).

(**) Activity observed by the STEREO/EUVI and SDO/AIA instruments.

(***) UT in year 2013. 

(\#) Frequency range of 16 MHz to 2.5 kHz for S/WAVES, from 14 MHz to 2.5 kHz for Wind/WAVES, and 14 to 47 MHz for BIRS.

(+) Letters a to e and numbers 1 to 3, as indicated in Fig. \ref{fig:Radio_waves}.

(oc.) Occultation due to relative position between spacecraft and AR.

(§) The AR6 location (represented by the CME apex direction), the CME eruption time, and flare size from this study.

(SA) Shock-accelerated type III radio burst (details given in main text).

 }
\end{table*}

\begin{figure*}
\centering
   \resizebox{\hsize}{!}{\includegraphics{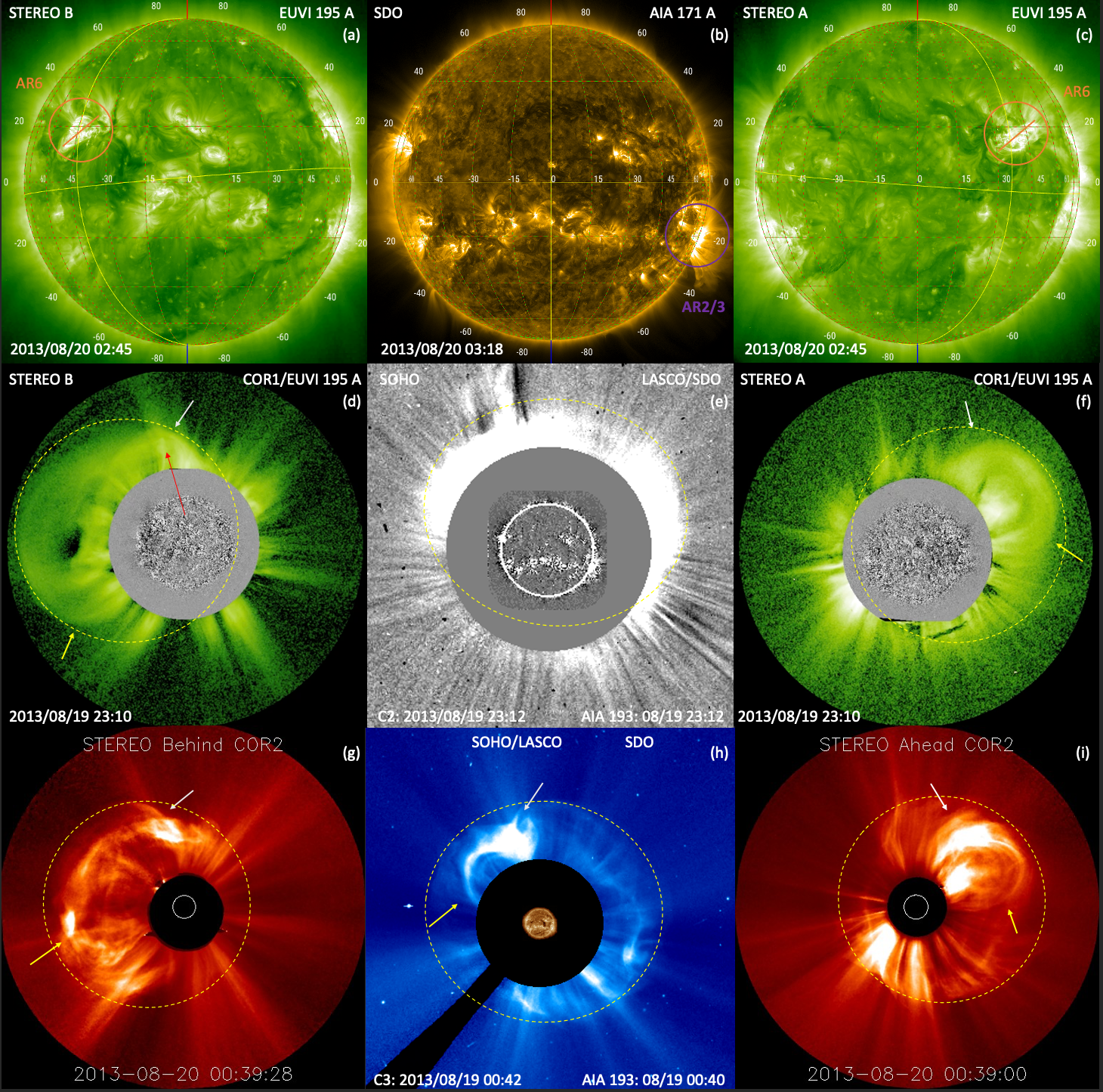}}
     \caption{EUV and coronagraph observations by STEREO-B (left), near-Earth (middle), and STEREO-A (right). The upper panels show EUV images taken by EUVI 195 {\AA}  and AIA 171 {\AA} (middle). AR6 is indicated by the orange circle in STEREO-A and STEREO-B, and AR2 and AR3  by the purple circle in SDO. The second and third rows show the evolution of the CME at two different times, seen from three different points of view, as observed by the LASCO (magnified image) and SECCHI coronagraphs. The CME-driven shock front is shown with yellow dashed curves (as a rough outline to guide the  eye) and the flux-rope structure is indicated with yellow and white arrows. The red arrow indicates the disturbance caused by the first stage (details given in main text).  Credit: \textit{JHelioviewer/cdaw.gsfc.nasa.gov}}
     \label{fig:EUV_COR}
\end{figure*}
\subsection{Energetic particle timing}
\label{sec:Analysis_VDA}

The velocity dispersion analysis (VDA) of an SEP event is based on determining the onset times of the event at a number of energies, and plotting these onset times as a function of the reciprocal of the particle velocities at respective energies. The slope indicates the effective path length (L) and the intercept gives the release time, assumed to be the same for all particles \citep[e.g.][]{Vainio2013}. It is known that the VDA is based on the assumption of a simultaneous release of particles of different energies at the Sun, a scatter-free propagation of the first arriving particles, and an accurate determination of onset times for the  in situ observations \citep{Zhao2019}. In this event only STEREO-A shows clear energy dispersion to perform VDA analysis (lower right panel in the upper part of Fig. \ref{fig:proton_electron_fluxes_solar_wind}). In the case of STEREO-B (lower left panel in the upper part of Fig. \ref{fig:proton_electron_fluxes_solar_wind}), as the intensity enhancement is very gradual, the onsets at different energies can only be calculated using 60-minute averaging, which makes the VDA analysis too ambiguous. For STEREO-A the onset time at different energies from SEPT (65-105 keV electrons) and HET (1.4-2.8 MeV electrons, and 13.6-15.1 MeV, 26.3-29.7 MeV, 60-100 MeV protons) were visually determined, zooming in around the onset time. Low-energy protons from the LET instrument (1.8-3.6 MeV, 4-6 MeV, 6-10 MeV protons) were not included in the VDA analysis as a preceding SEP background seems to mask the true onsets. The linear fit, shown in Fig. \ref{fig.VDA_STEREOA}, derives a VDA effective propagation path length of 3.55 $\pm$ 0.50 au, much longer than the length of $\sim$1.2 au expected for a nominal Parker spiral field and scatter-free propagation. It might indicate either a non-standard interplanetary magnetic field topology or that significant scattering effects are present in the propagation of the particles. It might also be possible that, in this event, the particles of different energies are not simultaneously released from the Sun. The time shifting analysis (TSA), which can be used for the first arriving particles, was also calculated based on \cite{Vainio2013}. We visually estimated the onset times of the near-relativistic electron channels for STEREO-A and MESSENGER, the 0.25-0.70 MeV electron channel for SOHO, and the 0.7-1.4 MeV electron channel for STEREO-B. Then we performed a back-shifting of the onset times to get the release times of the electrons at the Sun (shifted by light propagation to 1 au for comparison with electromagnetic observations). Table \ref{table_vda} summarizes the VDA and TSA results for different particle species and spacecraft. Column (2) presents the effective propagation path length for VDA and the ENLIL magnetic field line lengths used in TSA; there would be a difference of less than one minute for all spacecraft if Parker spiral lengths were used. Column (3) shows the estimated particle release time based on VDA and TSA analysis for STEREO-A, and on TSA analysis for STEREO-B, MESSENGER, and SOHO. We note that the MESSENGER onset time should be interpreted as the upper limit since the instrument field of view was pointing to the anti-sunward direction, and the first particles detected by the anti-sunward-looking instrument must have been scattered back towards the Sun from beyond the spacecraft location. Column (4) presents the shock heliospheric height at the release time using the coronal shock reconstruction, calculated at the connecting-to-observer-point \citep[cobpoint,][]{Sanahuja1995} for STEREO-A and MESSENGER, and at the shock nose for STEREO-B and SOHO, which is respectively not yet and never connected to the shock (Sect. \ref{sec:Modelling_GCS_Olmedo}). Column (6) presents the angle $\theta$\textsubscript{Bn} between the shock normal and the Parker field lines at the cobpoint.

Thus, the SEP estimated solar release time, shifted by light propagation time to 1 au, is $\sim$23:22 UT. As explained above, we note that this time should only be considered as an upper limit. At this time the heliocentric height of the quasi-parallel coronal shock is $\sim$6 R\textsubscript{$\odot$}, based on the Parker spiral field line.

\section{SEP parent solar source: Remote-sensing observations and data analysis}
\label{sec:OBSER_remote-sensing}

In order to identify the common solar origin for the particle enhancement observed at different spacecraft locations, the time of relevant remote-sensing observations ranges from 2013 August 19 at 20:30 UT to August 20 at 04:00 UT. This interval encompasses about three hours before the first electron onset observed at MESSENGER and the last proton onset at L1.  
A compilation of videos of the remote-sensing observations referenced in this section are available in the supplementary material of the online version of the paper.

\subsection{EUV and coronagraph observations}
\label{sec:EUV_Coronagraph}
The EUVI and AIA instruments captured low coronal signatures of CMEs in addition to other forms of solar activity, such as flares and jets. In these images we identified six main ARs as shown in Table \ref{table:Active_Regions}. Columns (1) to (4) show the AR number (from east to west longitudes), the AR location, the type of the activity, and the observation time, respectively. 
 \begin{figure}
   \resizebox{\hsize}{!}{\includegraphics{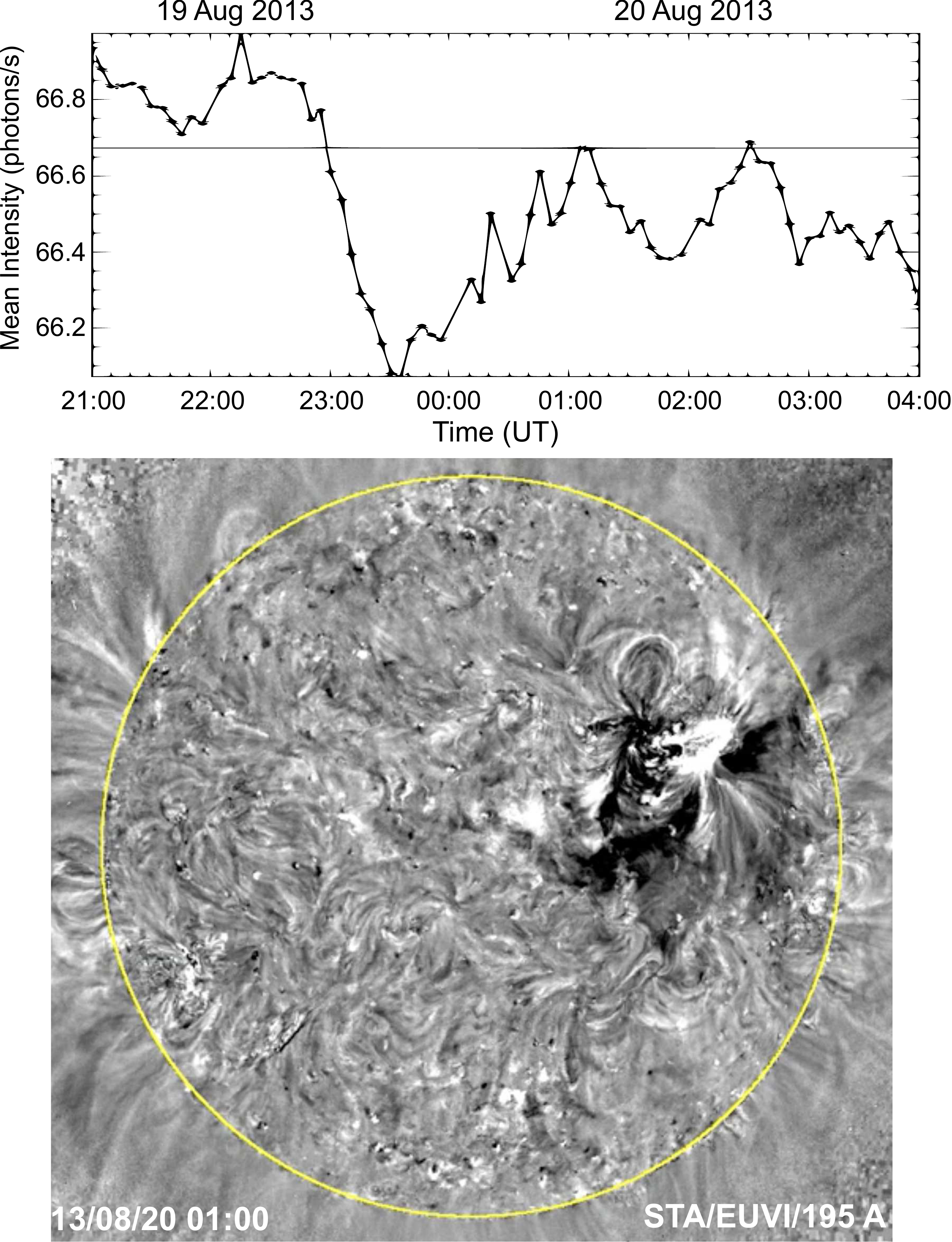}}
     \caption{EUVI-A observations and analysis. Top: STEREO-A light curve produced using full-disc EUVI 195 {\AA} images, where the horizontal line marks the mean intensity at 01:00 UT. Bottom: Snapshot at 01:00 UT from an EUVI-A 195 {\AA} percentage base-difference movie, where a brightening can be seen in AR6, near the CME dimming area. The yellow circle corresponds to 1 R\textsubscript{$\odot$}.}
     \label{fig:EUV_lightcurve}
\end{figure}
\begin{figure*}
\centering
  \resizebox{\hsize}{!}{\includegraphics{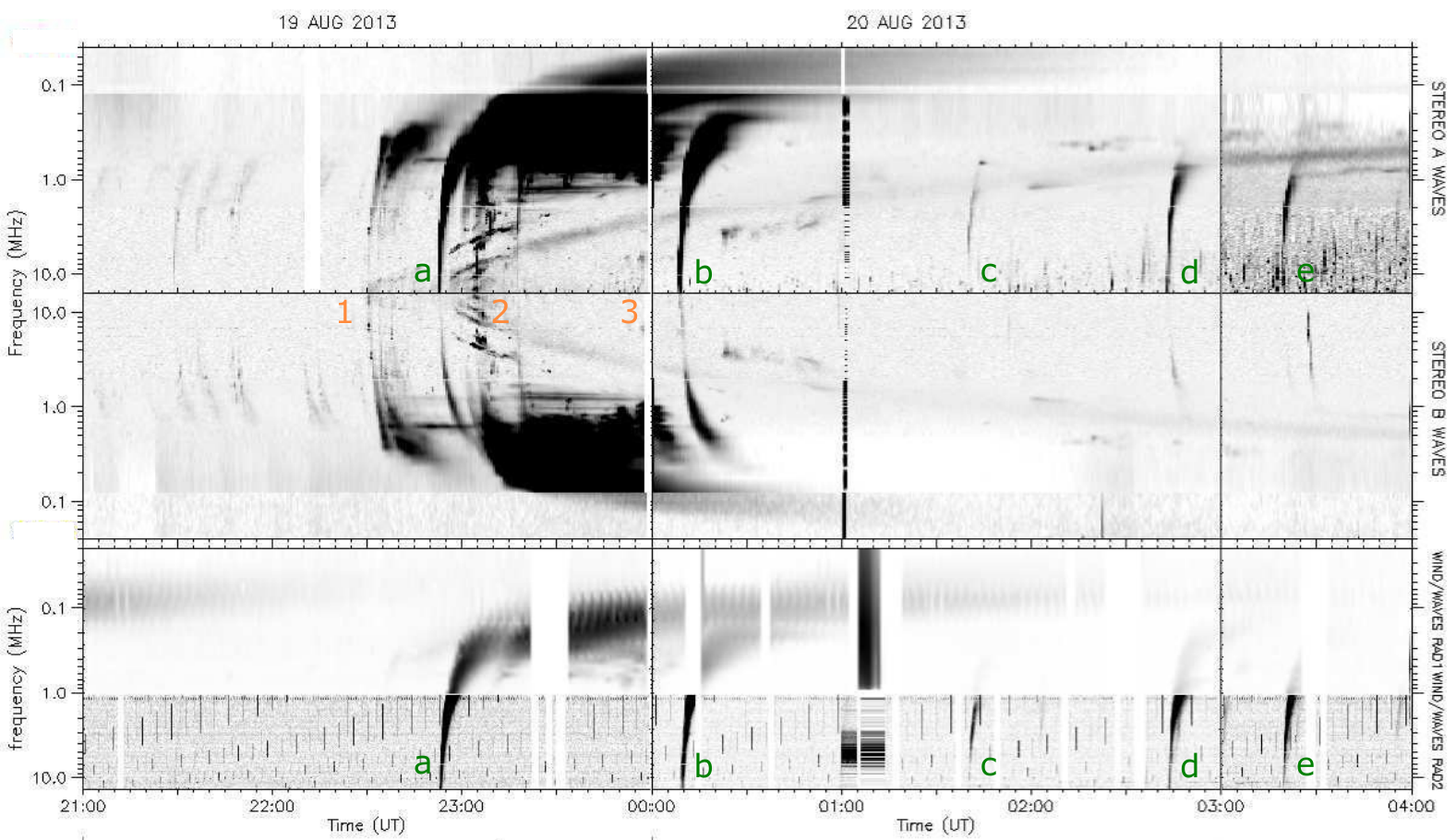}}
     \caption{Radio dynamic spectra observed by both S/WAVES and Wind/WAVES from 2013 August 19 at 21:00 UT to 2013 August 20 at 04:00 UT. Both S/WAVES present coronal--IP type II radio emission, with \textit{fp} and \textit{2fp} clearly present from $\sim$22:28 to $\sim$04:00 UT. Groups of (probably SA) type III bursts can be seen emerging from the type II, from $\sim$22:28 UT (labelled  with an orange `1') to $\sim$00:00 UT (orange `3'). Both type II and type III bursts are time related with the CME ejecting from AR6 (details given in main text). The orange number 2 indicates where \textit{fp} and \textit{2fp} could possibly be further broken down into more bands. S/WAVES-A and Wind/WAVES present flare-related type III bursts at different times (green letters a to e), caused by the activity in AR2, AR3, and AR5, and S/WAVES-B observes some of these flare-related type III bursts lower in frequency, due to occultation (Table \ref{table:Active_Regions}). Credit: \textit{Observatoire de Paris-Meudon}}
     \label{fig:Radio_waves}
\end{figure*}

We associate the SEP event with a shock driven by a CME ejecting from AR6, located near the far-side central meridian from Earth's perspective (last row in Table \ref{table:Active_Regions}), and observed as a two-stage eruption. The CME starting time given by the CDAW SOHO LASCO CME catalogue, calculated at 1 R\textsubscript{$\odot$} and referring to the second stage of the eruption, is 22:39 UT;  the first C2 appearance time is at 23:12 UT (Fig. \ref{fig:EUV_COR}(e)). The CME is described in the catalogue as a halo-type,  with an inferred CME speed of 877 km s$^{\ -1}$, and a positive acceleration of 4 km s$^{\ -2}$. The given CME mass and energy values are 1.6 $\times$ 10\textsuperscript{13} kg and 6.1 $\times$ 10\textsuperscript{25} J, respectively, but there is uncertainty due to a poor mass estimate as the mass measurement assumes that the CME material is in the sky-plane and the kinetic energy is obtained from the mass and linear speed \citep[][]{Vourlidas2000}. The CME mass and energy values are relatively large compared to overall CME averaged quantities \citep[Table 1 in ][]{Vourlidas2011}. Comparing the dynamical properties of CMEs and SEP intensities from statistical studies \citep[Table 1 in ][]{KahlerVourlidas2013}, the >50 MeV proton intensity enhancement of this SEP event agrees with the general trend inferred, where  higher masses and kinetic energies are associated with higher SEP intensities.

The first stage of the eruption is observed in EUVI-A 171 {\AA} base difference image at $\sim$20:35 UT as very faint and narrow moving loops (online material)\footnote{first\_stage\_lasco\_c2\_bd.mp4; first\_stage\_euvia\_195\_rd.mp4} that later extend in COR1-A and C2 fields of view, where it is first detected at $\sim$21:20 UT and $\sim$22:12 UT, respectively. Later on, the second stage structures become too dominant to clearly observe the first stage structures. The second stage resulted from a gradual rising motion of a closed coronal structure starting at $\sim$21:50 UT, more clearly seen in difference images, over the west limb from EUVI-A view. This motion suddenly accelerates at $\sim$22:10 UT, making the closed-loop structures move much more quickly. The structures rise over the STEREO-A north-west limb, seen in EUVI-A 195 {\AA} and 171 {\AA} running difference images, as moving loops continuing into COR1-A (online material)\footnote{first\_and\_second\_stage\_euvia\_195\_rd.mp4; first\_and\_second\_stage\_lasco\_c2\_bd.mp4}. The first stage structure effects can be seen as the distortion of the front, where the north part of the second and main stage is moving more slowly, as it is picking up the first flux-rope, as shown in Fig. \ref{fig:EUV_COR}(d), indicated with a red arrow. Afterwards, this interaction can also be observed as a brighter north part of the CME, indicating increased density, as seen in Fig. \ref{fig:EUV_COR}(g), \ref{fig:EUV_COR}(h), and \ref{fig:EUV_COR}(i), indicated with white arrows. Thus, the interaction of the two stages forms the CME, whose evolution is observed by COR1-B and COR2-B, C2 and C3, and COR1-A and COR2-A, as shown in the second and third rows of Fig. \ref{fig:EUV_COR}. During this early phase the CME appears as a flux-rope seen face-on by STEREO-B and edge-on by STEREO-A (outlined with the white and yellow arrows in Fig. \ref{fig:EUV_COR}(d) and \ref{fig:EUV_COR}(g), and in Fig. \ref{fig:EUV_COR}(f) and \ref{fig:EUV_COR}(i), respectively). This configuration is in agreement with the orientation of the post-eruptive arcades observed after the CME, shown as the orange line in Fig. \ref{fig:EUV_COR}(a) and Fig. \ref{fig:EUV_COR}(c). They show EUVI-B\&A 195 {\AA} images, respectively, a few hours after the CME eruption, where the AR6 is indicated with an orange circle. Although the eruption occurs on the far-side hemisphere from the Earth's perspective, SDO/AIA also observes the CME over the north and west limbs between $\sim$22:23 UT and $\sim$23:10 UT (online material)\footnote{second\_stage\_observed\_by\_sdo\_aia193\_pbd.mov}.

The evolution of the CME-driven shock can also be followed by COR1 and COR2, where the shock front is visible from $\sim$22:35 UT onwards. However, it is only at $\sim$22:55 UT when the CME-driven shock can be clearly observed for the first time from the three points of view, COR1-A, COR1-B, and C2 (online material)\footnote{shock\_evolution\_cor1a\_raw\_euvi195\_diff.mov; shock\_evolution\_cor1b\_raw\_euvi195\_diff.mov; first\_and\_second\_stage\_lasco\_c2\_bd.mp4}. The yellow dashed curves in the second and third rows of Fig. \ref{fig:EUV_COR} show the extent of the shock front later in time (as a rough outline to guide the  eye, so it does not represent any fitting results). Thus, the global geometry of this CME, formed by two stages, evolves as a single CME driving a single shock. 

Weak large-scale disturbances are seen to propagate across the disc in EUVI-A images during $\sim$22:30--22:50 UT (online material)\footnote{large\_scale\_weak\_disturbances\_euvia195\_pbd.mov}. However, they appear to be qualitatively different from what we know as EUV waves with a continuous well-defined semicircular front \citep[e.g.][]{Thompson1998}.

We produced EUVI light curves for both STEREO spacecraft, only shown for STEREO-A in the upper panel of Fig. \ref{fig:EUV_lightcurve}, which was computed using full-disc EUVI 195 {\AA} images. The small increase in intensity observed from $\sim$22:00 UT to $\sim$22:20 UT is mainly associated with the activity in the west limb from Earth's perspective (purple circle in Fig. \ref{fig:EUV_COR} (b)), corroborated by the light curve computed only for AR2/AR3 (not shown). From $\sim$22:20 UT to $\sim$23:40 UT the upper panel of Fig. \ref{fig:EUV_lightcurve} shows the dimming phase, observed as a decrease in brightness that follows the lifting of the CME. Before the dimming stage is finished, a flare starting at $\sim$23:30 UT causes an increase in the intensity of photons/s, peaking at 01:00 UT. The lower panel shows a snapshot from an EUVI-A 195 {\AA} percentage base-difference movie, at the time of the flare intensity peak, where the time of the initial image used is 19:00 UT on 2013 August 19. Thus, the EUVI light curves suggest that the flare represented a post-eruptive arcade (thermal phase) and lacked a clear impulsive phase. 

\subsection{X-ray observations}
Column (3) in Table \ref{table:Active_Regions} summarizes the type of solar activity (brightening or jet) observed during the period of study, and Col. (5) indicates the flare size if it was detected by X-ray Sensor (XRS) on board GOES\footnote{\url{ftp://ftp.swpc.noaa.gov/pub/warehouse/2013/2013_plots/xray/}}. XRS observes the Sun continuously from Earth's vantage point in two broadband soft X-ray channels. For the purposes of detecting the onset and the intensity of solar flares, it is usually sufficient to monitor the wavelength 1-8 {\AA} band. The lack of intense emissions observed during the period of study stems from the activity on the far side of the Sun from Earth's perspective. However, the GOES soft X-ray flux equivalent level of the activity observed on the far side, can be derived using the EUVI/195 {\AA} light curve as they are dominated by the Fe XXIV line at 192 {\AA} during flare times \citep{Nitta2013}. This process yields an equivalent GOES soft X-ray level of 3.3 $\times$ 10\textsuperscript{-5} W m\textsuperscript{-2} for STEREO-A and 3.1 $\times$ 10\textsuperscript{-5} W m\textsuperscript{-2} for STEREO-B. Due to the uncertainties involved in the process, the corresponding GOES soft X-ray level is calculated with an error of a factor of three \citep[scattering present in Fig. 7 in][]{Nitta2013}. Therefore, the equivalent GOES soft X-ray is within the range from M1 to M9, generally an M-class flare.

\begin{figure}[htbp] 
   \resizebox{\hsize}{!}{\includegraphics{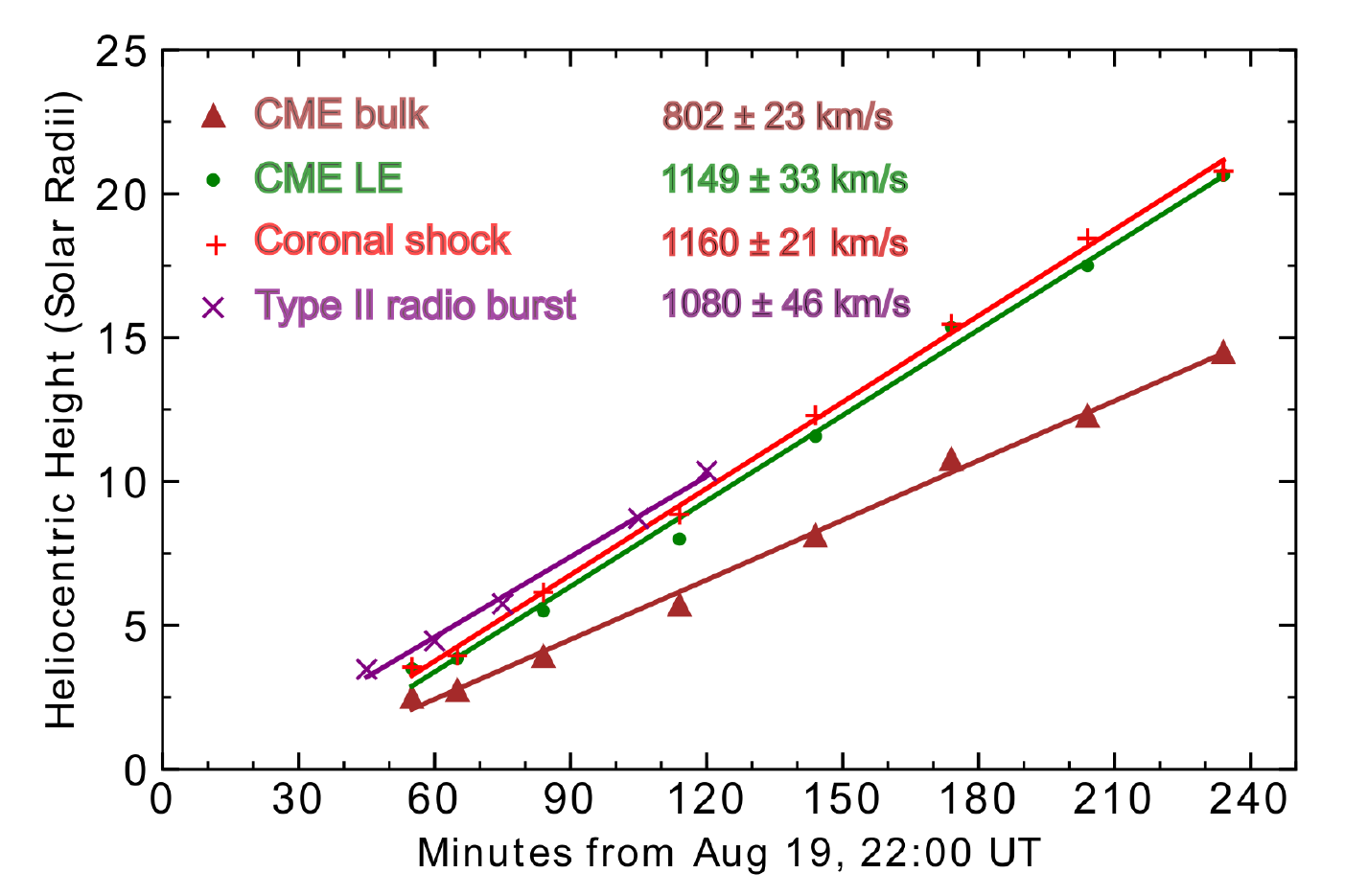}}
  \caption{Height evolution of the CME bulk and LE, coronal shock, and type II radio burst. The brown, green, and red lines show the CME (bulk and LE) and coronal shock height fits, respectively, from GCS and spheroid reconstruction. Type II radio burst drift, using the hybrid density model, is represented by the purple line (details given in main text). The legend shows the speeds corresponding to each linear fit.} 
    \label{fig:cme_shock_typeII}
\end{figure} 
\begin{figure*}
\centering
\resizebox{\hsize}{!}{\includegraphics{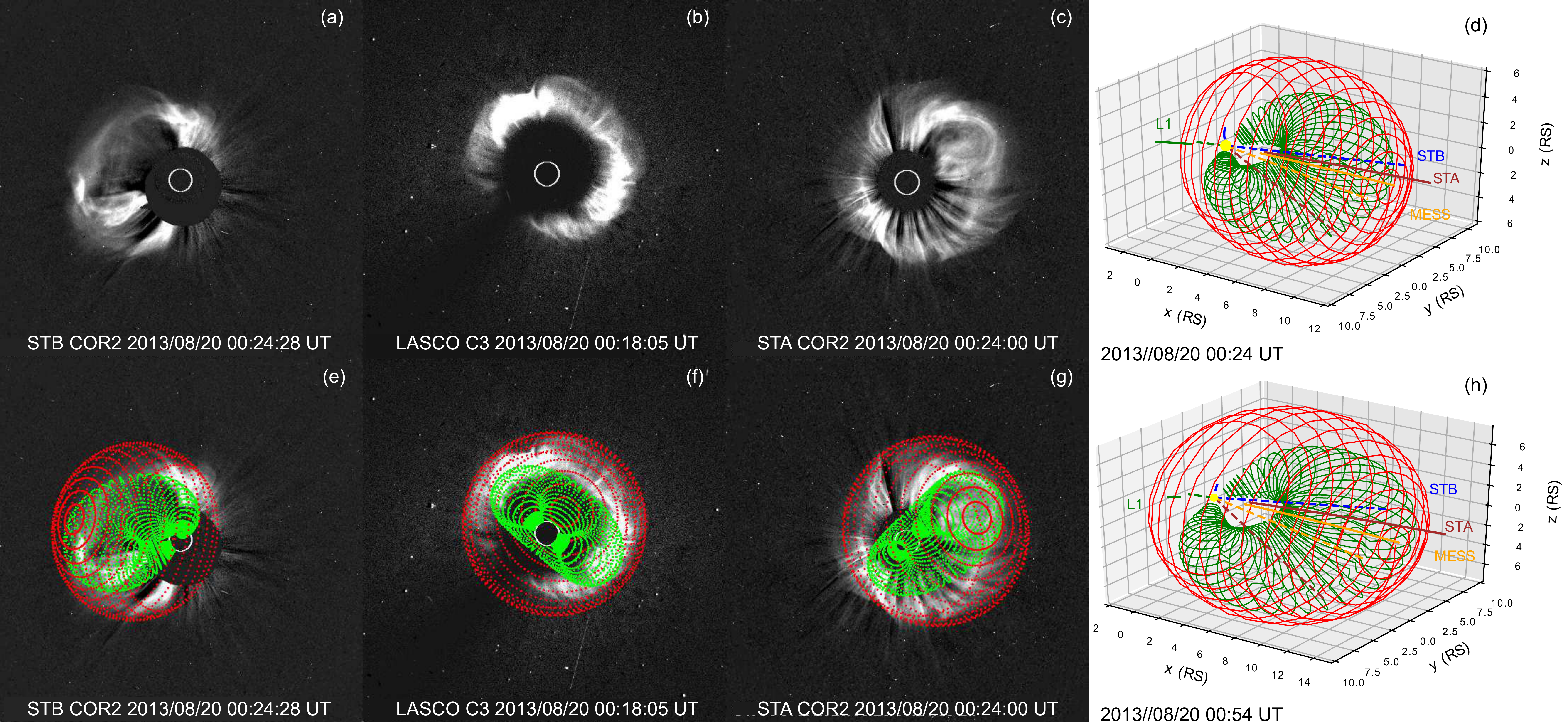}}
     \caption{Coronagraph images and GCS 3D reconstruction for the CME (green mesh) and associated driven shock (red mesh) as seen via three different instruments: COR2-B ((a),(e)), C3 ((b),(f)) and COR2-A ((c),(g)). The C3 images are magnified for  better visualization. The right plots ((d), (h)) show the 3D representation, in HEEQ reference in R\textsubscript{$\odot$}, of the CME and the driven shock, together with the nominal Parker spirals (solid lines) and the lines connecting the Sun (dashed lines), for STEREO-A (STA in red), STEREO-B (STB in blue), MESSENGER (MESS in orange), and the Earth (L1 in green). The Sun is shown as a yellow circle (not to scale).}
     \label{fig:GCS3D}
\end{figure*}

\subsection{Radio observations}
\label{sec:radio}
 
Radio observations provide insight in the coronal and IP processes that might be present related with the particle acceleration. Columns (6) to (9) in Table \ref{table:Active_Regions} present a summary of the different radio signatures observed by the S/WAVES, Wind/WAVES, and BIRS instruments during the period of study. The high- and low-frequency radio emission observed by the S/WAVES instruments (from 16 MHz to 2.5 kHz) on STEREO A and B and the Wind/WAVES experiment (from 14 MHz to 2.5 kHz) are shown in Fig. \ref{fig:Radio_waves}. Both S/WAVES present similar and complex radio emission, with intense activity during the period of observation. A first type II radio-emission is seen at $\sim$22:28 UT at 16 MHz (number 1 in Fig. \ref{fig:Radio_waves}), drifting slowly to 300 kHz at $\sim$04:00 UT. The type II emission could have started earlier, but there is no information above 16 MHz. These type II bursts, due to their low-frequency emission \citep[][and references therein]{Pohjolainen2007}, might be identified as coronal--IP type II radio bursts, associated with the shock driven by the CME, observed in COR1 at $\sim$22:35 UT. The IP type II emission persists for $\sim$8 hours (it can be tracked up to 06:00 UT, not shown). This time is longer that the average duration of $\sim$3-4 hours derived from the >50 MeV SEP statistical study performed by \cite{Kouloumvakos2019}, where in some extreme events the type II emission lasts for $\sim$12 hours. Assuming that the 16 MHz emission is related with the fundamental (\textit{fp}), and using the hybrid density model \citep[][details given below]{Vrsnak2004}, this frequency corresponds to a heliocentric height of $\sim$2.5 R\textsubscript{$\odot$} for the burst driver in the corona. A probable second harmonic, \textit{2fp}, might be present from $\sim$22:55 UT (at 16 MHz) to $\sim$04:00 UT (at 600 kHz), and at 23:00 UT (at $\sim$10 MHz, number 2 in Fig. \ref{fig:Radio_waves}) the \textit{fp} and \textit{2fp} are possibly further broken down into more bands \citep[][]{Pohjolainen2007}. Also at $\sim$23:00 UT the type II emission is more prominent starting at $\sim$5 MHz, which corresponds to a heliocentric height of $\sim$4.5 R\textsubscript{$\odot$}. The CME-driven shock speed can be estimated analysing the frequency drift. However, the heights and speeds inferred from radio burst dynamic spectra need to be handled with care since they  strongly depend on the chosen electron density model. To estimate the CME speed from radio observations, we assumed that the type II burst emission is formed by accelerated electrons, near the shock, at the leading-edge (LE) of the CME, although we can expect a certain offset between the burst driver and the shock \citep{Jebaraj2020}. The selection of a density model includes the knowledge of several conditions \citep[][]{Pohjolainen2007}. In this case, the hybrid model \citep{Vrsnak2004} is chosen because it more closely fits  the shock height-time evolution as observed by COR1-A and COR2-A. The hybrid model is a mixture of the \cite{Saito1970} and the \cite{Leblanc1998} models, with some small changes, and it can be used for connecting bursts in the corona and the IP space. Following the hybrid model, and knowing the frequency of the type II emission at a given time, we derived the heliocentric height of the burst driver in solar radii. Figure \ref{fig:cme_shock_typeII} shows that the frequency drift of the fundamental yields a linear fit speed of 1080 $\pm$ 46 km s\textsuperscript{-1} (purple line), which is the same for both S/WAVES within the given uncertainty. The points were taken from 22:45 UT (8.5 MHz, i.e. 3.2 R\textsubscript{$\odot$}) up to 00:00 UT (1 MHz, corresponding to 10.4 R\textsubscript{$\odot$}),  when the lowest frequency available from the hybrid model was reached. Based on the type II radio burst fit, the derived height of the CME-driven shock at the estimated particle solar release time is 6.6 $\pm$ 0.5 R\textsubscript{$\odot$}.

Wind/WAVES presents, simultaneously with S/WAVES, flare-related type III bursts from at least 16 MHz to lower frequencies at different times (letters a to e in Fig. \ref{fig:Radio_waves}), as summarized in Table \ref{table:Active_Regions}, that are not related with the SEP event. In particular, the activity on the west limb from Earth's perspective associated with the anisotropy period observed at Wind (Sect. \ref{sec:Anisotr}) might be related to the type III radio burst labelled `e'. Some of these type III radio bursts start higher in frequency, as they can be seen in the BIRS experiment, from 14 to 47 MHz (Col. 9 in Table \ref{table:Active_Regions}), and in Learmonth spectrograph, from 25 to 180 MHz (except type III at 22:55 UT). Both S/WAVES present groups of type III emission bursting for several minutes, and covering 16 MHz to lower than 100 kHz in frequency, from $\sim$22:28 UT to $\sim$00:00 UT (from number 1 to number 3 in Fig. \ref{fig:Radio_waves}). This type III emission is relatively intense at low (hectometric) frequencies, and it is enhanced as observed by S/WAVES-A at around
23:00 UT (number 2 in Fig. \ref{fig:Radio_waves}). It seems that the type III bursts are emerging from the type II emission, but there is no radio information above 16 MHz to confirm it. These type III bursts can be observed down to 300 kHz (first group) and even lower (the other groups), which suggests that electrons are being lost to the IP medium through open magnetic field lines, as normally happens in energetic events \citep{Cane1981}. These type III bursts could be identified as shock-accelerated type III bursts showing particular properties: their timing is not associated with any starting or peaking flare (light curve in Fig. \ref{fig:EUV_lightcurve}), and they seem to be produced from type II bursts.
The corresponding SEP event shows proton increases of at least 50 MeV, consistent with the previous studies about the relationship between radio emission and SEP events; for example \citet{Cane2002} concluded that >20 MeV proton increases were always accompanied by type III or type III-l bursts. The M-class flare seen on AR6, starting at $\sim$23:30 UT and with a brightening peak at $\sim$01:00 UT, is not associated with any type III burst, suggesting that there are no accelerated electron beams escaping from the flare region.

\subsection{CME 3D reconstruction}
\label{sec:Modelling_GCS}

To characterize the CME associated with the SEP event, mainly in terms of CME speed, width, and location, we took advantage of the multi-view spacecraft observations and reconstructed the 3D CME using the GCS model. As described in \cite{Thernisien2006GCS} and \cite{Thernisien2011}, the GCS model uses the geometry of what looks like a hollow croissant to fit a flux-rope structure using coronagraph images from multiple viewpoints. The GCS model was developed to integrate both the self-similar expansion and the flux-rope three-dimensional morphology, using the following constraints: the legs of the structure are conical, the front is pseudo-circular, the cross section is circular, and it expands in a self-similar way. GCS estimates the CME position, direction, 3D extent, 3D speed, and the CME orientation. The sensitivity (deviations) in the parameters of the GCS analysis is given in Table 2 of \cite{Thernisien2009}. The COR1/2-A and COR1/2-B, and C2 and C3 quasi-simultaneous images were used to fit the flux-rope shape of CME at different times, where COR1 and C2, and COR2 and C3 data were used at lower and higher altitudes, respectively, to help track  the propagation and evolution of the CME more precisely. The routine used for the reconstruction is \textit{rtcloudwidget.pro}, available as part of the \textit{scraytrace} package in the SolarSoft IDL library\footnote{\url{http://www.lmsal.com/solarsoft/}}. 

Figure \ref{fig:GCS3D} shows the coronagraph images (upper left) and GCS fit analysis (lower left) for the CME, on August 20 at 00:24 UT, where the green mesh represent the flux-rope structure erupting from AR6, seen from the points of view of  STEREO-B (e), SOHO (f), and STEREO-A (g). Figure \ref{fig:GCS3D}(g) reveals that the CME is more oriented towards STEREO-A  than to STEREO-B, as the green wire-frame is more symmetric with respect to the centre than in STEREO-B image (e). Figures \ref{fig:GCS3D}(d) and \ref{fig:GCS3D}(h) show the 3D reconstruction of the flux-rope structure at two different times, and its relative position to the Sun-spacecraft direction for STEREO-A, MESSENGER, STEREO-B, and L1. 

The CME 3D reconstruction, from $\sim$3.5 to $\sim$21 R\textsubscript{$\odot$}, presents a smooth behaviour with minor variations in the values of its longitude ($\sim$+171$^{\circ}$) and tilt ($\sim$36$^{\circ}$), which is the angle relative to the solar equator. However, the structure seems to experience a slight southward deflection, from a CME apex of +8$^{\circ}$ at 23:25 UT, corresponding to $\sim$6 R\textsubscript{$\odot$}, to -5$^{\circ}$ at $\sim$ 00:54 UT, corresponding to $\sim$15 R\textsubscript{$\odot$}. The aspect ratio ($\sim$0.4) remains almost constant at all times, following the hypothesis of self-similar expansion, and so does the half-angle ($\sim$59$^{\circ}$).

The linear fit of the height evolution in time for the CME bulk and LE are shown in Fig. \ref{fig:cme_shock_typeII}. The calculated bulk speed (i.e. speed of the centre of the flux-rope at the apex) is 802 $\pm$ 23 km s\textsuperscript{-1}, while the LE speed is 1149 $\pm$ 33  km s\textsuperscript{-1}. The width of the CME was estimated based on \cite{Dumbovic2019}, where the angular extent in the equatorial plane is represented by ${R\textsubscript{maj}-{(R\textsubscript{maj}-R\textsubscript{min})} \times |tilt|/90}$, obtaining a result of 121$^{\circ}$. The value of $R\textsubscript{maj}$ (face-on CME half-width) was calculated adding $R\textsubscript{min}$ (edge-on CME half-width) to the half-angle, and $R\textsubscript{min}$ was calculated as the $\arcsin(ratio)$. The CME width deviation was derived from the mean half-angle error, estimated by \cite{Thernisien2009} as +13$^{\circ}$/-7$^{\circ}$.
Thus, the CME is ejecting from W171N08 and it is wide ($\sim$121$^{\circ}$) and fast ($\sim$1149 km s\textsuperscript{-1}).

 \begin{table*}[ht]
\caption{First intersection between the coronal shock and magnetic field lines connecting the spacecraft\textsuperscript{a}}
\label{table:olmedo_parker_ENLIL}

\begin{tabular}{ccccccccc}
\hline
\hline
s/c&  \multicolumn{2}{c}{Estimated first}& \multicolumn{2}{c}{Shock height from} &\multicolumn{2}{c}{$\theta$\textsubscript{Bn} at the} &\multicolumn{2}{c}{Cobpoint}\\
&\multicolumn{2}{c}{intersection time (UT)\textsuperscript{b}}& \multicolumn{2}{c}{Sun centre (R\textsubscript{$\odot$})} &\multicolumn{2}{c}{cobpoint (deg)\textsuperscript{c}}&&\\& Parker&ENLIL & Parker&ENLIL  & Parker&ENLIL &  Parker&ENLIL\\
(1)&(2)&(3)&(4)&(5)&(6)&(7)&(8)&(9)\\
 \hline
(STA & 22:41 $\pm$ 10 min&22:42 $\pm$ 10 min  & 2.6 $\pm$ 0.3&2.8 $\pm$ 0.3  & 37 $\pm$ 2&24 $\pm$ 2 &E157S04&E164S05)\\
(MESS & 22:38 $\pm$ 10 min&22:42 $\pm$ 10 min  & 2.5 $\pm$ 0.3&3.0 $\pm$ 0.3  & 5 $\pm$ 1&9 $\pm$ 1&W179N01&W174N01) \\
STA & 22:55 $\pm$ 10 min&22:55 $\pm$ 10 min  & 3.3 $\pm$ 0.3&3.2 $\pm$ 0.3  & 17 $\pm$ 2&22 $\pm$ 2 &E157S04&E164S05\\
MESS & 22:55 $\pm$ 10 min&22:55 $\pm$ 10 min  & 3.5 $\pm$ 0.3&3.5 $\pm$ 0.3  & 4 $\pm$ 1&7 $\pm$ 1&W179N01&W174N01 \\

\hline
\\
\end{tabular}

\footnotesize{ \textbf{Notes.}

\textsuperscript{a} As the estimated first intersection times occur in the first step of the coronal shock reconstruction using coronagraph data from both STEREO and LASCO (last two lines), we also present the first intersection results calculated using EUVI/COR1 data from both STEREO instruments (first two lines in brackets). Details given in main text.

\textsuperscript{b} Time on 2013 August 19.

\textsuperscript{c} Angle between the shock normal and the upstream magnetic field line at the cobpoint. The given uncertainties are rounded to one degree. They are based on the shock height deviation due to the coronagraph images cadence (the fluctuations in the magnetic field lines, due to waves and turbulence, or in the shock front geometry are not considered). 

}
\end{table*}
\subsection{Coronal shock 3D reconstruction}
\label{sec:Modelling_GCS_Olmedo}
In order to gain a detailed understanding of the magnetic connectivity to the CME-driven shock associated with the SEP event, the coronal shock 3D reconstruction was performed using the model developed by \cite{Olmedo2013}. This model fits the coronal shock to an spheroid shape, although the real wave probably presents some differences from the ideal contour \citep{Susino2015}. Some examples of the application of the model can be found in \cite{Makela2015} and \cite{Xie2017}. The model used here for the shock reconstruction is similar to that applied by \cite{Kwon2014}, who used an ellipsoid shape for the fitting. As there is no reference in the literature that describes the spheroid model, we include the details of the process here. The cartesian equation for a spheroid that is rotationally symmetric along the z-axis is given by $(x^2+y^2)/a^2+z^2/c^2=1$, where the two distinct semi-axes are denoted $a$ in the azimuthal direction and $c$ in the radial direction. The geometry of the spheroid is defined by three parameters: the height ($h$) of the spheroid in the radial direction from the solar centre in units of solar radii (i.e. the height of the nose of the spheroid in the z-axis), the self-similarity coefficient ($\kappa$), and the ellipticity ($e$). The semi-axis ($a$) of the spheroid in the azimuthal direction is related with the parameter $\kappa$ such that $a=(h-1)\times~\kappa$. For a given height ($h$), and knowing the self-similarity parameter ($\kappa$), which takes values between 0.01 and 2, the semi-axis ($a$) in the azimuthal direction is determined. The ellipticity ($e$) is  analogous to the eccentricity, and if the ellipticity is known, the radial semi-axis ($c$) can be calculated.

The model uses a spheroid shape to fit the CME-driven shock using quasi-simultaneous images from COR1 and COR2, and from C2 and C3. The images underwent a basic process for calibration, and base-difference or running-difference procedure was used to highlight the front of the shock better. By varying the ellipticity ($e$), which is always positive, the relationship between $c$ and $a$ can be changed. For $c>a$ a prolate spheroid is defined, and an oblate spheroid for $c<a$. In the routine, however, the difference between an oblate or prolate spheroid is implemented by choosing a negative or positive value of the ellipticity ($e$), respectively. To determine the latitude and longitude coordinates of the source region, we calculated the intersection point of the radial axis (z-axis) of the spheroid and the solar surface. The routine used for the reconstruction is also \textit{rtcloudwidget.pro} (reference in Sect. \ref{sec:Modelling_GCS}). 

The main shock reconstruction period, using the three vantage points of view, covered from 22:55 UT on August 19 when the shock height was $\sim$3.5 R\textsubscript{$\odot$}, to 01:54 UT on August 20, corresponding to a shock height of $\sim$21 R\textsubscript{$\odot$}. 
Signatures of the shock formation are observed earlier in EUVI and COR1 images, from $\sim$22:25 UT and $\sim$22:35 UT onwards, respectively, when no observations from the Earth's perspective were available. Using these data we   also performed an earlier 3D reconstruction of the CME-driven shock. However, as the shock front is not as clear as from $\sim$22:55 UT onwards, we considered that the shock reconstruction using EUV images for the period prior to this time was less reliable. Therefore, we based our further analysis of the 3D shock reconstruction on white-light coronagraph data from three viewpoints. 

The CME-driven shock 3D reconstruction at 00:24 UT on August 20 is shown in Fig. \ref{fig:GCS3D}. The red mesh seen overplotted on the coronagraphs images represents the 2D projections of the 3D reconstruction of the coronal shock, as seen from COR2-B (e), C2 and C3 (f), and COR2-A (g). The 3D view of the CME-driven shock and the nominal Parker magnetic field lines, starting at 2.5 R\textsubscript{$\odot$}, for STEREO-A, STEREO-B, Earth and MESSENGER, are shown in Fig. \ref{fig:GCS3D}(d) and \ref{fig:GCS3D}(h). We note that the rear part of the spheroid was removed, exactly in the plane containing the source origin of the shock at the Sun. It is remarkable that the nose of the CME-driven shock is oriented towards MESSENGER and STEREO-A, and the magnetic field lines connecting STEREO-A and MESSENGER intersect the spheroid representing the CME-driven shock. 

The 3D reconstructed shock parameters are consistent during the main reconstruction period. The resultant spheroid is oblate ($e$=0.46) and the self-similarity coefficient ($\kappa$) is $\sim$0.69. The longitude and latitude values show that the origin at the Sun of the coronal shock is located at W173N02, slightly west and south from AR6 (W171N08). Lastly, the coronal shock speed, estimated as the linear fit of the evolution of the shock height, is 1160 $\pm$ 21 km s\textsuperscript{-1} (red line in Fig. \ref{fig:cme_shock_typeII}). We note that this shock speed is very similar to the estimated type II drift (Sect. \ref{sec:radio} and purple line in Fig. \ref{fig:cme_shock_typeII}). Based on the spheroid model analysis, the derived height of the CME-driven shock at the estimated particle solar release time is 6.0 $\pm$ 0.5 R \textsubscript{$\odot$}. 

Table \ref{table:olmedo_parker_ENLIL} shows the (analytically calculated) intersection parameters between the spheroid and the nominal Parker magnetic field lines connecting to each spacecraft. The first two lines in brackets correspond to the earlier 3D shock reconstruction mentioned above. Column (2) shows the first time the reconstructed CME-driven shock intersects the magnetic field lines connecting the spacecraft. Column (4) represents the coronal shock heliospheric height at the intersection time. Column (6) presents $\theta$\textsubscript{Bn}, which is  the value of the angle between the shock normal and the upstream magnetic field line at the cobpoint. Column (8) shows the cobpoint coordinates. Thus, the first connection to the shock, for STEREO-A (STA) and MESSENGER (MESS), is at least $\sim$27 minutes before the particle solar release time at 23:22 $\pm$ 8 min. Taking into account the earlier 3D reconstruction, the first connection to the shock for STEREO-A and MESSENGER would be $\sim$22:41 UT and $\sim$22:38 UT, respectively, which is $\sim$44 minutes before the particle solar release time. Based on the shock reconstruction up to $\sim$21 R\textsubscript{$\odot$}, STEREO-B and L1 are not connected to the coronal shock at any time. The heliocentric shock height near the shock nose at the first intersection time is $\sim$2.5 R\textsubscript{$\odot$}, in comparison with the $\sim$6 R\textsubscript{$\odot$} at the estimated particle release time. Both MESSENGER and STEREO-A, connecting to the CME front area, are observing quasi-parallel shocks. The MESSENGER cobpoint is located near the shock nose and the STEREO-A cobpoint is longitudinally separated less than $\sim$30$^{\circ}$ from the CME-driven shock nose.

\section{Heliospheric conditions: In situ solar wind observations and ENLIL simulation}
\label{sec:heliospheric_conditions}
\subsection{In situ plasma and magnetic field observations}
\label{sec:OBSER_Solar_wind}
The heliospheric conditions at the time of the particle release, in which the particles propagate, can significantly affect the SEP timing and intensity profiles. Multi-point solar wind and IMF observations provide a comprehensive understanding of the geometry, not only of the IP structures and their possible influence in the propagation of the SEPs, but also of the shocks and their role in forming the intensity-time profiles. 

As presented below, according to in situ solar wind and magnetic field data, an IP shock and ICME was observed at MESSENGER, and both STEREO spacecraft, consistent with the two-stage CME erupting from AR6, as no other CMEs within a few hours were ejected from the same AR and surrounding region. The detailed analysis of the IP shock and ICME observed by MESSENGER, at 0.33 au, and by STEREO-A and STEREO-B, at $\sim$1 au, is the focus of a second study \citep{Rodriguez-Garcia2021CME}. Because the Earth  is located on the opposite side of the Sun from AR6, it is exceedingly unlikely that the reported ICME transients from August 22 to August 24 on the near-Earth ICME list by Richardson and Cane are associated with the SEP parent CME. 

In situ plasma and magnetic field observations by PLASTIC and the Magnetic Field Experiment on board STEREO, and by SWEPAM, SWICS (proton density), and the Magnetic Field Experiment on board ACE are shown in the lower part of Fig. \ref{fig:proton_electron_fluxes_solar_wind}, where the different IP structures observed by the spacecraft were classified using the lists summarized in Sect. \ref{sec:INSTRUM}.

The left panels of bottom part of Fig. \ref{fig:proton_electron_fluxes_solar_wind} show that at the time of the relativistic electron onset (arrow in the speed panel), STEREO-B is embedded in a slow solar wind stream (V\textsubscript{sw} $\sim$ 320 km s\textsuperscript{-1}). A short while later, during the rising phase of the SEP event, STEREO-B observes a SIR, corresponding to the salmon shaded area, where the stream interface is represented by the grey dash-dotted line. The arrival of the IP shock, at 02:10 UT on August 22, is indicated with a vertical black line. STEREO-B, located 53{$^{\circ}$} west of CME apex direction (estimated at the time of the ICME arrival), shows a gradual onset, with increased intensities as the spacecraft becomes connected to the shock. Maximum intensity, for <10 MeV protons, is observed when the spacecraft encounters magnetic field lines that are connected to the shock when it is beyond 1 AU, with an exponential decay following. This peak observed after the shock can be related with the spacecraft connecting to a stronger region of the shock closer to the nose \citep{Cane1988}, although it has little influence on the near-relativistic electrons and protons >30 MeV. Thus, STEREO-B and also STEREO-A (discussed below) match the expected behaviour for a gradual SEP event seen from two different solar longitudes relative to the CME-driven shock \citep{Cane1988}. This behaviour is also consistent with the observed anisotropies (Sect. \ref{sec:Anisotr}). The ICME arriving at STEREO-B is shown as  a grey area and seems to have little influence in the SEP profile. The solar wind after the ICME is a fast wind stream with Alfvénic-like fluctuations in the magnetic field, as some correlation is observed between the magnetic field and solar wind speed in the normal (N) component from RTN coordinate system.

The right panels of Fig. \ref{fig:proton_electron_fluxes_solar_wind} show that at the time of the relativistic electron onset, the solar wind speed at STEREO-A is $\sim$400 km s\textsuperscript{-1}. The IP shock arrives at 07:05 UT on August 22. STEREO-A, located 24{$^{\circ}$} east of the CME apex direction (estimated at the time of the ICME arrival), observes a peak in intensity for < 6 MeV ions when the shock arrives, followed by a decay. The local shock has little influence on the near-relativistic electrons (although this is difficult to evaluate, due to proton contamination around the shock transit time), and for > 10 MeV. A sudden decrease in protons up to 30 MeV is observed at the time of the ICME arrival, at 23:15 UT on August 22, followed by a transient depression around 12 hours later. The decrease observed in near-relativistic electrons at the same times is unsure due to proton contamination.

The central panels of Fig.  \ref{fig:proton_electron_fluxes_solar_wind} show that at the time of the relativistic electron onset the solar wind speed at L1 is $\sim$450 km s\textsuperscript{-1}. In a later phase of the SEP event, not affecting the onset, the near-Earth spacecraft are observing several IP structures not related with the activity in AR6, located close to the far-side central meridian from Earth's perspective. Just before the beginning of August 21, a first IP shock (vertical dashed line), and the corresponding ICME (first green shaded area), embedded in a SIR \citep[salmon shaded area, based on][]{Chi2018SIRcatalog} arrive at the L1 spacecraft. The perturbed IP magnetic field lines may be affecting the SEP time profile, with a possible additional contribution of in situ particle acceleration (protons up to several MeV). In particular, the two small peaks observed at the beginning and ending of August 21 could be related to the local compression, observed as increased magnetic field, density, and speed. This behaviour is also consistent with the observed anisotropies (Sect. \ref{sec:Anisotr}). A second ICME (second dashed line and second green shaded area)   arrives before the end of August 22 with little influence on the SEP profile. The last shock arrives on August 24, but no significant effect is observed on the SEP time profile either. 

\begin{figure}[htbp] 
   \resizebox{\hsize}{!}{\includegraphics{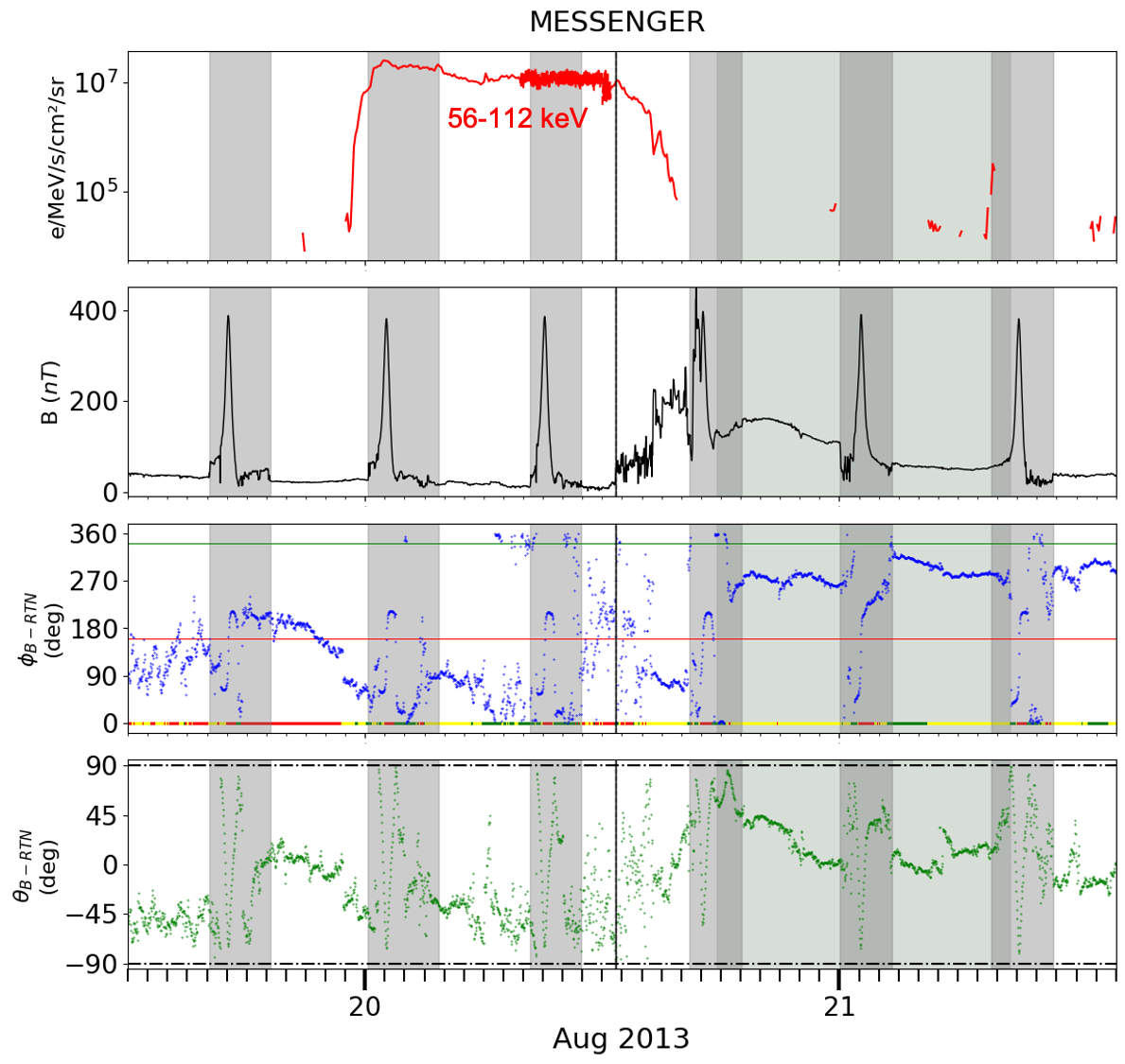}}
     \caption{MESSENGER energetic electron and magnetic field observations. The panels show from top to bottom: Near-relativistic electron intensities, magnetic field magnitude, and magnetic field angles $\phi$\textsubscript{B-RTN} and $\theta$\textsubscript{B-RTN}. 
     The coloured lines and lower band in the $\phi$\textsubscript{B-RTN} angle panel as in Fig. \ref{fig:proton_electron_fluxes_solar_wind}. The grey shaded areas delimit the intervals related with Mercury's magnetosphere. An IP shock and ICME arriving at MESSENGER are indicated by the black vertical line and the green shaded area, respectively.}
     \label{fig:messenger}
\end{figure}
The energetic particle and magnetic field observations by MESSENGER, in orbit around Mercury and located 6{$^{\circ}$} east of the CME apex direction (estimated at the time of the ICME arrival) are shown in Fig. \ref{fig:messenger}. The grey shaded areas delimit the intervals related with Mercury's magnetosphere, identified by eye, while outside these intervals the spacecraft was in the solar wind. We observe a sharp decrease in the SEP profile starting with the arrival of the IP shock transit on August 20 at 12:40 UT (vertical black line). The influence of the ICME (green shaded area) arriving on August 20 at 17:49 UT, with a change in the magnetic field polarity, and ending on August 21 at 08:38 UT, cannot be evaluated due to the lack of particle data during this period. 

\subsection{ENLIL model}
\label{sec:Modelling_ENLIL}
The  WSA-ENLIL + Cone model \citep{Odstrcil2004} is a global 3D MHD model\footnote{\url{https://ccmc.gsfc.nasa.gov/models/modelinfo.php?model=ENLIL\%20with\%20Cone\%20Model}} that provides a time-dependent background characterization of the heliosphere outside of 21.5 R\textsubscript{$\odot$}. ENLIL uses time-dependent GONG magnetic field data as a basis, into which spherical shaped high-pressure structures without any internal magnetic field are inserted to mimic observed CME-associated solar wind disturbances. The ENLIL-modelled CME has an artificially higher pressure to compensate for the lack of a strong magnetic field. At the CME nose, the dynamic pressure dominates and the compression is strongest, thus the ENLIL+Cone model can often simulate the CME nose well, despite the lack of internal magnetic field for the CME driver. In contrast, at the flank or edge of a CME, the interaction of CME magnetic field with background solar wind cannot be ignored anymore, so the lack of internal magnetic field would influence the simulation of CME flanks. This affects the shock parameters such as the shock obliquity derived from the model \citep{Bain2016}. The reliability of ensemble CME-arrival predictions depends strongly on the initial CME input parameters, such as speed, direction, and width \citep{Mays2015,Kay2020}, but also on the errors that can arise in the ambient-model parameters and on the accuracy of the solar-wind background derived from coronal maps. Based on \cite{Wold2018}, the mean absolute arrival-time prediction error is 10.4 $\pm$ 0.9 hours, with a tendency to an early prediction error of -4.0 hours.

\begin{figure}
\centering
   \resizebox{\hsize}{!}{\includegraphics{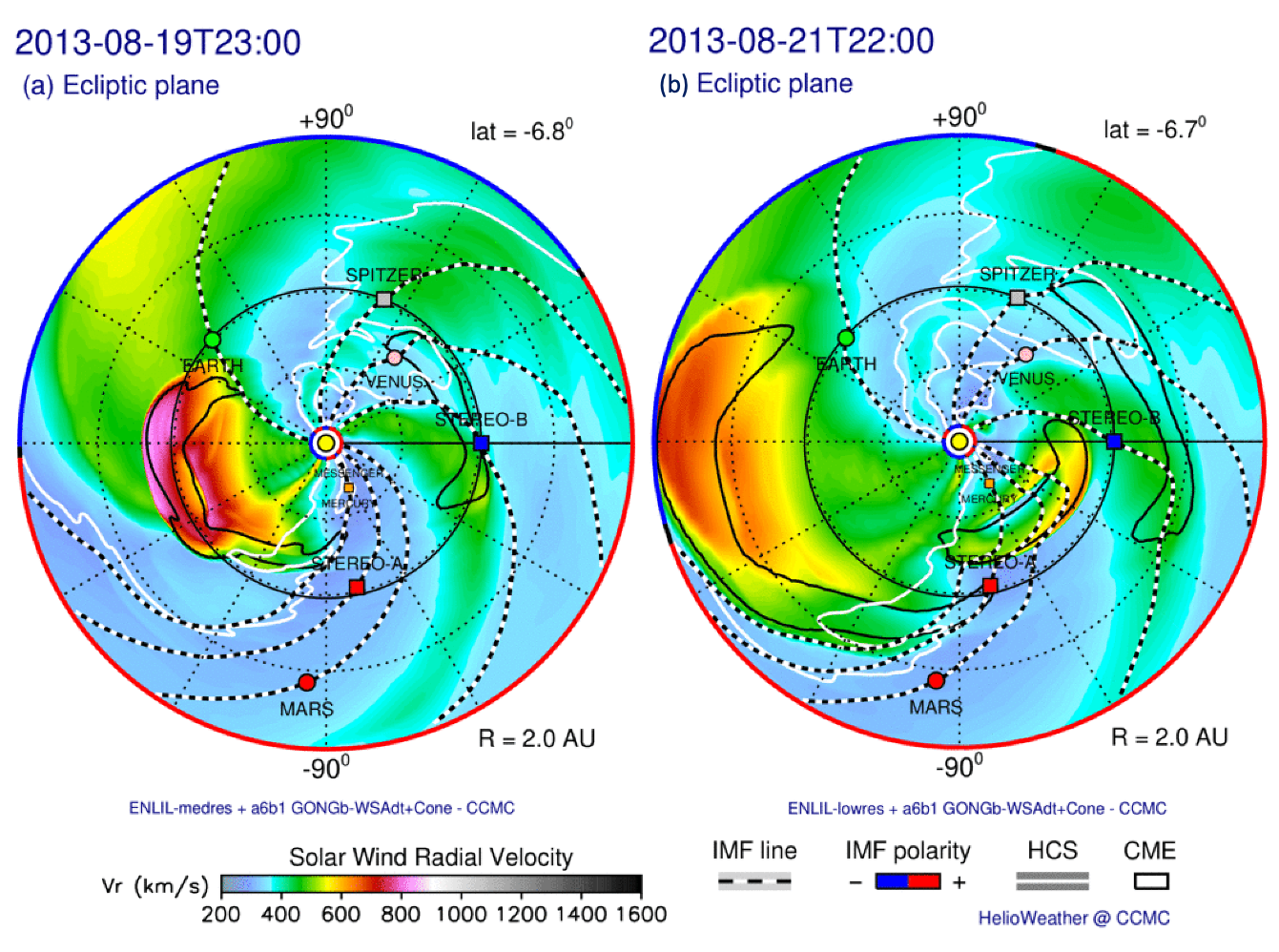}}
     \caption{Radial velocity contour plots from the ENLIL simulation in the ecliptic plane. The black and white dashed lines represent the IMF lines and the black contours track the ICMEs. The white lines correspond to the HCS, which separates the regions with opposite magnetic polarity, shown in blue (negative) or red (positive) on the outer edge of the simulation region. Left panel (a): Magnetic connectivity of the different spacecraft around the particle solar release time. Right panel (b): Magnetic field line connecting STEREO-B intersects for the first time the IP shock driven by the ICME. Credit: \textit{CCMC}}
     \label{fig:ENLIL_onset_arrival}
\end{figure}
 \begin{table*}[htbp]
\centering
\caption{Time line of the 2013 August 19 SEP event.}
\label{table:SEPTtimeLine}
\begin{tabularx}{1\textwidth}{ccccccc} 
\hline
\hline
Date  & Delay to  & Event   & Mission/ & This study/  & Notes\\ 
(UT)&CME (hr)\textsuperscript{a}&description&instrument&reference&\\
 \hline
19/08 \hspace{1mm} 20:35 & -1:35 & CME starts (1\textsubscript{st} stage) & STA/EUVI &Sect. \ref{sec:EUV_Coronagraph} & AR6 (W171N08) \\
19/08 \hspace{1mm} 22:10 & 0:00 & CME starts (2\textsubscript{nd} stage) & STA/EUVI &Sect. \ref{sec:EUV_Coronagraph} & AR6 (W171N08) \\
19/08 \hspace{1mm} 22:28 & 0:18 & Type II bursts (\textit{fp}) &STA \& STB/WAVES &  1 in Fig. \ref{fig:Radio_waves} &$\sim$16 MHz/2.5 R\textsubscript{$\odot$} \\
19/08 \hspace{1mm} 22:28 & 0:18 & SA type III bursts &STA \& STB/WAVES &  1 in Fig. \ref{fig:Radio_waves} & First group  \\
19/08 \hspace{1mm} 22:35 & 0:25 & Coronal shock in COR1  &STA/SECCHI & Sect. \ref{sec:Modelling_GCS_Olmedo} & W173N02\\
19/08 \hspace{1mm} 22:42 & 0:32 & First shock connection  &STA\& MESS & Table \ref{table:olmedo_parker_ENLIL} & EUVI/ENLIL\\
19/08 \hspace{1mm} 23:00 & 0:50 & SA type III bursts &STA \& STB/WAVES &  2 in Fig. \ref{fig:Radio_waves} & Second group  \\
19/08 \hspace{1mm} 23:19 & 1:09 & 71-112 keV e onset   &MESSENGER/EPS & Fig. \ref{fig:SEP_SC_Position}(c) & Anti-Sun /$\pm$5 min \\
19/08 \hspace{1mm} 23:22 & 1:12 & 71-112 keV e SRT\textsuperscript{b}  &MESSENGER/EPS & TSA (time shifted) & Anti-Sun /$\pm$5 min \\
19/08 \hspace{1mm} 23:30 & 1:20 & $\sim$M soft-X ray starts & STA/EUVI & Light curve in Fig. \ref{fig:EUV_lightcurve} & $\pm$5 min  \\
20/08 \hspace{1mm} 00:15 & 2:05 & 0.7-1.4 MeV e onset  &STA/HET & Fig. \ref{fig:proton_electron_fluxes_solar_wind} (top right panel) & $\pm$15 min \\
20/08 \hspace{1mm} 00:15 & 2:05 & 60-100 MeV p onset  &STA/HET & Fig. \ref{fig:proton_electron_fluxes_solar_wind} (top right panel) & $\pm$15 min \\
20/08 \hspace{1mm} 00:25 & 2:15 & 65-105 keV e onset  &STA/SEPT & Fig. \ref{fig:proton_electron_fluxes_solar_wind} (top right panel) & $\pm$10 min \\
20/08 \hspace{1mm} 01:00 & 2:50 & $\sim$M soft-X ray peaks & STA/EUVI & Light curve in Fig. \ref{fig:EUV_lightcurve} & $\pm$5 min   \\
20/08 \hspace{1mm} 01:00 & 2:50 & 0.25-0.70 MeV e onset &SOHO/EPHIN & Fig. \ref{fig:proton_electron_fluxes_solar_wind} (top central panel) & $\pm$60 min  \\
20/08 \hspace{1mm} 03:00 & 4:50 & 0.7-1.4 MeV e onset  &STB/HET & Fig. \ref{fig:proton_electron_fluxes_solar_wind} (top left panel) & $\pm$60 min \\
20/08 \hspace{1mm} 03:00 & 4:50 & 25-50 MeV p onset &SOHO/EPHIN & Fig. \ref{fig:proton_electron_fluxes_solar_wind} (bottom central panel) & $\pm$10 min \\
20/08 \hspace{1mm} 04:00 & 5:50 & >27 MeV p onset &MARS ODY./HEND & Fig. \ref{fig:SEP_SC_Position}(b) & $\pm$20 min \\
20/08 \hspace{1mm} 06:00 & 7:50 & $\sim$1.5 days anisotr. period &STA/LET & Fig. \ref{fig:anisotropies_protons} (right panels) & 1.8-3.6 MeV p \\
20/08 \hspace{1mm} 06:45 & 8:35 & 60-100 MeV p onset  &STB/HET & Fig. \ref{fig:proton_electron_fluxes_solar_wind} (bottom left panel) & $\pm$15 min \\
20/08 \hspace{1mm} 12:41 & 14:31 & IP shock arrival  &MESSENGER & ICME  catalogue & 975 km/s\textsuperscript{c} \\
20/08 \hspace{1mm} 17:49 & 17:39 & ICME starts  &MESSENGER & In situ IMF data & 860 km/s\textsuperscript{c} \\
21/08 \hspace{1mm} 05:00 & 30:50 & $\sim$1.5 days anisotr. period &STB/LET & Fig. \ref{fig:anisotropies_protons} (right panels) & 1.8-3.6 MeV p \\
21/08 \hspace{1mm} 08:38 & 34:28 & ICME ends  &MESSENGER & In situ IMF data & - \\
21/08 \hspace{1mm} 22:01 & 47:51 & First shock connection  &STEREO-B & ENLIL model & - \\
22/08 \hspace{1mm} 02:09 & 51:59 & IP shock arrival  &STB & IP Helsinki DB & 615 km/s\textsuperscript{d} \\
22/08 \hspace{1mm} 07:05 & 56:55 & IP shock arrival  &STA & IP Helsinki DB & 483 km/s\textsuperscript{d} \\
22/08 \hspace{1mm} 23:15 & 73:05 & ICME starts &STA & ICME catalogue & 475 km/s\textsuperscript{d} \\

23/08 \hspace{1mm} 07:00 & 80:50 & ICME ends  &STB & ICME catalogue & - \\
24/08 \hspace{1mm} 23:25 & 97:15 & ICME ends  &STA & ICME catalogue &- \\
 \hline
\\
\end{tabularx}
\begin{flushleft}
\footnotesize{\textbf{Notes.}

\textsuperscript{a} The second and main stage of CME is taken for time reference.

\textsuperscript{b} Solar release time (upper limit), shifted 8.3 min for comparison with electromagnetic observations from 1 au.

\textsuperscript{c} Average transient speed from the Sun to spacecraft location. 

\textsuperscript{d} In situ speed of the transient.

}
\end{flushleft}
\end{table*}

The magnetic connectivity at the onset time can be relevant to the understanding of the SEP observations, and considering the ENLIL-modelled varying solar wind conditions to calculate the IMF lines is an alternative to using the nominal Parker spirals. The preconditioning of the heliosphere and the interaction of the IP structures that might be present at the onset time can actively influence this connectivity. Therefore, the ENLIL simulation time ranges from August 16 to August 26 (i.e.  from four days before to six days after the SEP event). This interval encompasses the possible previous CME and CME-driven shocks that may influence the particle propagation at the onset time, and the ICME evolution through the IP medium up to 2.1 au. For this purpose, the GCS 3D reconstruction process presented in Sect. \ref{sec:Modelling_GCS} was also used for the five relevant prior CMEs erupting in the time range of August 16 to August 20. For this simulation, we used the CME LE parameters (position and speed) rather than the bulk (bright core, if present) as they often capture the overall and great impact of the high-pressure structures better. For the study of this particular scenario, we   performed two simulations with two different grid resolutions (low and medium), but otherwise identical runs. The  medium-resolution run (786x60x180 for 2.1 au radius, $\pm$60{$^{\circ}$} latitude, 360{$^{\circ}$} longitude) seems to increase the difficulty of the model in deriving the ICME flank geometry, observed by STEREO-B. The derived shock parameters from the medium-resolution grid fluctuate significantly, while those from the low-resolution grid (384x30x90 for 2.1 au radius, $\pm$60{$^{\circ}$} latitude, 360{$^{\circ}$} longitude) match the in situ data more closely. The CME and model set-up parameters, and the results of the simulations are available on the Community Coordinated Modeling Center (CCMC) website\footnote{\url{https://ccmc.gsfc.nasa.gov/database_SH/Laura_Rodriguez-Garcia_093020_SH_1.php} (low resolution)\label{footnote low resolution}} \footnote{\url{https://ccmc.gsfc.nasa.gov/database_SH/Laura_Rodriguez-Garcia_093020_SH_2.php} (medium resolution)\label{footnote med resolution}}.

The information derived from the modelled IMF lines at the SEP onset time, using the medium-resolution run, which provides the information about the magnetic field lines with smaller uncertainty, is summarized in Table \ref{table:Magnetic footpoint location} and represented in the left panel of Figure \ref{fig:ENLIL_onset_arrival}. As the magnetic footpoints and the IMF lines passing through the different spacecraft are given by the model, the length of the magnetic field lines can be calculated. For each spacecraft, Cols. (5), (6), (9), and (10) of Table \ref{table:Magnetic footpoint location} show, respectively, the solar wind speed, the magnetic field line length, and the longitude and latitude of the footpoint locations on August 19 at 23:00 UT. The ENLIL lines are computed up to 21.5 R\textsubscript{$\odot$}, and radially extended to 5 R\textsubscript{$\odot$}, which corresponds to the height of the shock at the estimated particle released time ($\sim$23:22 UT), based on the MESSENGER TSA analysis presented in Table \ref{table_vda}. At this height (5 R\textsubscript{$\odot$}), the shock is simulated as a quasi-parallel shock in the corona based on the spheroid model and ENLIL magnetic field lines ($\theta\textsubscript{Bn}$= 8{$^{\circ}$} in Col. (7) in Table \ref{table_vda}). 

Based  on the connection angle shown in Col. (11), STEREO-B and Earth seems to have respectively a better and worse magnetic connection to the source area (W171N08) from the ENLIL model than they would have from the nominal Parker spiral. Column (6) shows that the ENLIL-modelled and Parker spiral magnetic field lines lengths are comparable for all spacecraft. Figure \ref{fig:ENLIL_onset_arrival} (a) shows a snapshot of the ENLIL simulation at the onset time, where the black contours track the ICMEs, as they are set as regions when the density ratio to the background is higher than a certain threshold. This figure reveals that there are IP structures present near STEREO-B and Earth that might be modifying the connectivity at these locations with respect to nominal Parker spirals. A region with increased solar wind speed might arrive at STEREO-B at around the SEP onset time, which could correspond to the SIR observed in Fig. \ref{fig:proton_electron_fluxes_solar_wind}. STEREO-A might be separated from MESSENGER by an HCS crossing. 

Table \ref{table:olmedo_parker_ENLIL} presents the parameters of the intersection between the reconstructed coronal shock and ENLIL magnetic field lines (extending radially from 21.5 R\textsubscript{$\odot$} to the Sun centre). It shows that the MESSENGER and STEREO-A first connection to the shock takes place at $\sim$22:55 UT when the shock is at respectively $\sim$3.5 and $\sim$3.2 R\textsubscript{$\odot$} (Cols. (3) and (5)). Thus, the first connection to the shock occurs at least $\sim$27 minutes before the estimated particle solar release time at $\sim$23:22 UT. If the earlier 3D reconstruction is considered (Sect. \ref{sec:Modelling_GCS_Olmedo}), this first connection time for MESSENGER and STEREO-A would be at $\sim$22:42 UT,  $\sim$40 minutes before the estimated particle solar release time. The MESSENGER cobpoint is located very close to the shock nose and the STEREO-A cobpoint is longitudinally separated by $\sim$23$^{\circ}$ east from the CME-driven shock nose (Col. (7)). 

As STEREO-B is not connected to the shock in the corona, the ENLIL simulated time-dependent CME-driven shock connectivity outside 21.5 R\textsubscript{$\odot$} might help to understand the SEP intensity profile at STEREO-B. The location of the simulated shock is identified as speed increases by more than 20 km/s compared to the ambient simulation along magnetic field lines connected to the observer. To derive a particular shock position in the shock files, a background run without that particular CME is subtracted from the run with that particular CME \citep[][]{Bain2016}. For example, to get the position of the shock associated with CME 1, the run of background solar wind is subtracted from the run with background solar wind and CME 1. To get the shock position associated with CME 2, the run of background solar wind and CME 1 is subtracted from the run with background solar wind, CME 1, and CME 2. A general limitation of the model regarding the shock simulation is that the CME input parameters are derived from the fit of the CME LE, but the shock itself may be wider \citep[][]{Bain2016}. Based on the low-resolution run\textsuperscript{\ref{footnote low resolution}}, STEREO-B becomes magnetically connected to the shock for the first time at a radial distance of 0.63 au, on August 21 at 22:01 UT, which corresponds to the first non-zero numbers in the shock parameters file from the simulation. Figure \ref{fig:ENLIL_onset_arrival} (b) shows the configuration of heliosphere at this time. Thus, STEREO-B's first connection to the shock occurs within a few hours of the time at which STEREO/LET observed an increase in protons at pitch angle 0$^{\circ}$ (left panel in Fig. \ref{fig:anisotropies_protons}), which starts around August 21 at 05:00 UT. Due to the positive magnetic field polarity of this period, the particle increase observed at pitch angle 0$^{\circ}$ corresponds to particles propagating away from the Sun along the IMF lines.

\section{Summary and discussion}
\label{sec:discussion}
On 2013 August 19, a large widespread SEP event was observed by STEREO-A, STEREO-B, the L1 spacecraft, MESSENGER, and Mars Odyssey, spanning a longitudinal range of 222$^{\circ}$ in the ecliptic plane, where >2 MeV electron and >50 MeV proton intensity enhancements were observed by the first three spacecraft. The single solar AR (named AR6) associated with the particle enhancements observed in all these spacecraft, from where a CME was ejected, was located near the far-side central meridian from Earth's perspective. The SEP intensity profiles seen by MESSENGER and STEREO-A were very different, although the longitudinal separation between the spacecraft was only 15$^{\circ}$, and the footpoint longitudinal separations to AR6 were $\Delta$$\phi$=3$^{\circ}$ and $\Delta$$\phi$=25$^{\circ}$, respectively. MESSENGER observed a prompt steep rise, and STEREO-A observed a delayed and gradual increase. STEREO-B and L1, poorly connected to AR6 ($\Delta$$\phi$=102$^{\circ}$ and $\Delta$$\phi$=-138$^{\circ}$, respectively) presented  time profiles similar to those of STEREO-A, but with more gradual and delayed flux increases. Table \ref{table:SEPTtimeLine} summarizes the timeline for the SEP event and the signatures of the associated solar eruption. The wide ($\sim$121$^{\circ}$) and fast ($\sim$1149 km s\textsuperscript{-1}) CME associated with this unusual widespread SEP event was ejected in two stages. The second and main eruption, which occurred at $\sim$22:10 UT, is taken as the time reference for the timeline. 

The CME-driven shock was observed for the first time at $\sim$22:35 UT in COR1, with a resulting speed of $\sim$1160 km s\textsuperscript{-1}, based on the spheroid model. S/WAVES observed type II radio bursts starting at $\sim$22:28 UT that persisted for $\sim$8 hours, and whose frequency drift yields a linear fit speed of 1080 km s\textsuperscript{-1}, very similar to the 3D reconstructed CME-driven shock speed. Groups of type III bursts were observed from $\sim$22:28 to $\sim$00:00 UT, extending to low frequencies, which suggests that electrons are being lost to the IP medium through open magnetic field lines. This type III emission was relatively intense at low frequencies and was more prominent around $\sim$23:00 UT (second group of emissions), when the type II bursts were more intense too. These type III bursts seem to originate from the coronal--IP type II emission (although there is no higher-frequency information to confirm it) and they might only be associated with the presence of the shock (SA type III). We note that these groups of type III radio emission are similar to what \cite{Cane2002} called type III-l, which could be related with a signature of flare-related acceleration following the departure of the CME. However, the scenario where the type III bursts are shock-accelerated seems to be reasonable in this unusual SEP event. This is based on the timing of the type III emissions, starting at the same time of the type II emissions and 60 minutes (first group) to 30 minutes (second group) before the post-eruptive arcade started (with no type III emission related with the flare, as detailed below), the long electron rise times observed by SEPT, the hard spectral indices in the electron spectra observed by both STEREO, and the presence of MeV electrons (as shown in Fig. \ref{fig:proton_electron_fluxes_solar_wind}). This scenario is corroborated by the statistical study carried by \cite{Dresing2020}, where the authors propose that electron events with the above characteristics cannot be explained by a pure flare scenario. These events require  another acceleration mechanism that involves a prolonged particle injection. Thus, the CME-driven shock might play an important role in accelerating the electrons observed in this SEP event.

Accompanying the CME eruption, an M-class flare was seen as a post-eruptive arcade starting at $\sim$23:30 UT and with a peak of intensity at $\sim$01:00 UT. The EUVI light curves do not clearly reveal a flare impulsive phase, and there is no type III emission related to the flare, which suggests that there is little or no particle release or acceleration from the flare site. The M-class flare is too late to explain the particle acceleration in this unusual SEP event, as the estimated particle solar release (shifted) time is $\sim$23:22 UT. This result also reflects that of \cite{Kahler2000}, who provided evidence against the possibility of solar post-eruptive arcade contributions to gradual SEP events. We note that although there is only 8 minutes difference between the flare onset and solar particle release time, the latter is based on MESSENGER observations, and should be considered as an upper limit due to the anti-sunward direction pointing of the instrument. Thus, the observations support that the likely solar source associated with the widespread event is the shock driven by the CME. 

The first spacecraft to observe a particle enhancement was MESSENGER ($\Delta\phi$=3$^{\circ}$), already connected to the shock nose area before $\sim$22:42 UT (based on the earlier 3D shock reconstruction). The quasi-relativistic electron onset was observed at $\sim$23:27 UT (shifted by light propagation time to 1 au in order to compare with electromagnetic observations). The 59-minute delay with respect to the weak supercritical shock wave formation, as indicated by the type II burst \citep{Classen1998} observed at $\sim$22:28 UT and corresponding to a height of $\sim$2.5 R\textsubscript{$\odot$}, could be due to a combination of factors. Firstly, the connection of the spacecraft to the shock is estimated to occur later ($\sim$22:42 UT); secondly, the shock might not efficiently accelerate energetic particles until it is strong enough \citep{Kouloumvakos2019}. The last factor is  the anti-Sun pointing of MESSENGER instrument. At $\sim$23:22 UT, which corresponds to the estimated release time of particles (shifted to 1 au), the shock heliocentric height is $\sim$5 R\textsubscript{$\odot$} (using ENLIL simulated magnetic field lines). This height agrees (at the higher end of the range) with the statistical study on shock properties associated with >50 MeV proton SEP events during solar cycle 24, carried out by \cite{Kouloumvakos2019}. They conclude that shock supercriticality is reached between $\sim$1.2 and $\sim$6 R\textsubscript{$\odot$}, and the solar particle release times tend to occur from a few minutes up to $\sim$80 minutes after the shock waves become supercritical for well-connected spacecraft ($\Delta\phi$ <70$^{\circ}$). Thus, in this scenario, where the shock is the main source of the particle acceleration, the time between the first and second group of type III bursts could be related to the time the shock needs to reach supercriticality. We note that at the time of the second group of type III bursts, the emission is enhanced and coincides with more prominent type II bursts starting at $\sim$5 MHz, which corresponds to a heliocentric height of $\sim$4.5 R\textsubscript{$\odot$}.

The second spacecraft to observe an increase in particles was STEREO-A ($\Delta\phi$=25$^{\circ}$), already connected to an area $\sim$23$^{\circ}$ east from the shock nose, before $\sim$22:42 UT (based on the earlier 3D shock reconstruction). The onset of the relativistic electron (0.7-1.4 MeV) was estimated at $\sim$00:15 UT, which corresponds to a 107-minute delay with respect to the type II--first group of SA type III radio bursts. \cite{Richardson2014} examined the connection-angle dependence of electron and proton delays, derived from observed particle onset times during solar cycle 24 from STEREO and near-Earth spacecraft, relative to the onset of the associated type III radio emissions. If we compare the 107-minute delay with the \cite{Richardson2014} study, we find that STEREO-A relativistic electron onset delay is unusually longer. We note that both MESSENGER and STEREO-A are simultaneously connected to the shock at $\sim$22:42 UT, but at two different shock areas, when the shock is at $\sim$3.0 R\textsubscript{$\odot$} (based on the earlier 3D shock reconstruction and ENLIL model). Thus, assuming that the CME-driven shock is the main source of the particle acceleration, our results indicate that the unusual delay observed in STEREO-A electron onset could be the result of three factors (the first two  closely related to each other): firstly, the time the shock needs to become supercritical (as explained above for MESSENGER); secondly, the difference in the cobpoint location between MESSENGER and STEREO-A, whose cobpoint is separated $\sim$23$^{\circ}$ east from the nose area; and thirdly, significant particle scattering between the locations of Mercury, at 0.33 au, and STEREO-A, at 0.97 au (in addition to the possible delay due to the larger heliocentric distance of STEREO-A). This is based on the prompt and sharp particle enhancement observed by MESSENGER, the delayed and gradual particle flux increase observed by STEREO-A, the low anisotropies observed by STEREO-A, and the long effective path length derived from STEREO-A velocity dispersion analysis. If we associated the second group of type III bursts to the SEP event, considered  the time the shock would start accelerating energetic particles, the delay would be reduced to 74 minutes. However, this delay would be still considered longer in comparison with the statistical study of \cite{Richardson2014}, probably due to the scattering present in the particle transport to STEREO-A location. 

The last spacecraft to observe an enhancement in particles are the L1 spacecraft and STEREO-B. SOHO ($\Delta\phi$=-138$^{\circ}$), observing lower intensities for all energies and species than STEREO-B ($\Delta\phi$=102$^{\circ}$), seems to present earlier onsets for relativistic electrons than STEREO-B. The gradual increase, which can influence the onset determination (given with 60-minute uncertainty for both locations), and the different instrument designs could be behind the earlier onsets observed at SOHO. Although SOHO/EPHIN observed a higher preceding SEP background at some energies than STEREO-B, SOHO/EPHIN has a much larger geometric factor and a wider field of view. It is also equipped with a lateral anti-coincidence, resulting in much lower instrumental background than STEREO/HET (Fig. 3 in \citealt{Richardson2014}) and better statistics. STEREO-B relativistic (0.7-1.4 MeV) electron flux started to rise at $\sim$03:00 UT when the spacecraft was not yet connected to the shock along the IP magnetic field lines (based on the 3D shock reconstruction), which corresponds to 272-minute delay from the type II--first group of SA type III radio bursts. This could mean either that the propagating coronal shock kinematics were not well modelled (although there is good agreement between type II radio burst drift and 3D modelled shock height evolution, as presented in Fig. \ref{fig:cme_shock_typeII}), or that the cross-field transport, supported by the low anisotropies and delayed rising fluxes observed by STEREO-B, might be a key factor in the particle propagation, and there is no need to be magnetically connected to the shock in order to observe a SEP event \citep[e.g.][]{Dresing2012,Laitinen2013,Dresing2014,Droege2016,Kollhoff2021}. The L1 spacecraft, with a relativistic (0.25-0.70 MeV) electron onset at $\sim$01:00 UT (152-minute delay to the first group of type III bursts), observed an even more gradual increase. Because they were not connected at any time with the shock (based on the 3D shock reconstruction and ENLIL modelling), the low anisotropies and the slow and delayed rising fluxes suggest that the observations are most likely explained in terms of perpendicular diffusion \citep[e.g.][]{Qin2015,Strauss2017}.

The acceleration and propagation of particles in this SEP event were also influenced by the interplanetary structures present in the heliosphere. In particular, according to the in situ solar wind data and ENLIL modelling, an ICME was observed at MESSENGER and both STEREO, consistent with the two-stage CME erupting from AR6, as no other CMEs within a few hours were ejected from the same AR or surrounding region. We observed an abrupt decrease in intensity following the CME-driven shock arrival to MESSENGER. STEREO-B observed a late increase in proton anisotropies that occurs 17 hours ahead of the time at which ENLIL modelling predicts the first magnetic connection to the shock on August 21 at $\sim$22:01 UT, when the IP shock was at a heliocentric distance of $\sim$0.6 au. For both STEREO spacecraft the acceleration of particles continued for the <10 MeV protons during the evolution of the shock through the heliosphere until its unexpected earlier arrival to STEREO-B (August 22 at 02:09 UT) than to STEREO-A (August 22 at 07:05 UT). STEREO-A observed a particle intensity decrease after the shock arrival, while STEREO-B presented a particle flux increase some hours later, probably related with the spacecraft connecting to a stronger region of the shock closer to the nose. STEREO-A observed a sudden decrease in particle intensities at the time of the ICME arrival (August 22 at 23:15 UT), and also a transient cavity $\sim$12 hours later, probably related with the closed magnetic field topology of the IP structure. The SEP profile at STEREO-B, that encountered the ICME very close to a flank, seems not to be influenced by the ICME arrival. The detailed analysis of the evolution of the IP shock and ICME, observed at the locations of MESSENGER (0.33 au) and both STEREO spacecraft (1 au), including the evidence of a complex magnetic structure within the ICME, is the focus of a second study \citep{Rodriguez-Garcia2021CME}.

\section{Conclusions}
\label{sec:Conclusions}
Our main conclusions can be summarized as follows:
\begin{description}
 \item[$\bullet$ Solar origin:] The solar source associated with the widespread SEP event is likely the shock driven by the two-stage CME eruption observed near the far-side central meridian from Earth's perspective, as the M-class flare observed as a post-eruptive arcade is too late to explain the particle acceleration.
 
  \item[$\bullet$ Radio dynamic spectra:] Coronal--IP type II radio emission related with the presence of the CME-driven shock is prominent and persists for $\sim$8 hours. Groups of shock-accelerated type III bursts are observed from $\sim$50 minutes before to $\sim$40 minutes after the solar release time of the particles. This type III emission, enhanced around the time of the particle release time, is relatively intense at low (hectometric) frequency, and extends to low frequencies (lower than 100 kHz).     
 
 \item[$\bullet$ CME-driven coronal shock:] The SEP acceleration might start when the shock heliocentric height is $\sim$5 R\textsubscript{$\odot$}. As for other previously observed widespread SEP events, the longitudinal extent of the CME-driven shock does not explain by itself the wide spread of particles in the heliosphere as STEREO-B and L1 are not connected with the coronal shock when the respective particle fluxes start to rise. The shock longitudinal width, figured as to be the equivalent 2x\textit{R\textsubscript{maj}}, is 116$^{\circ}$ at 21.5 R\textsubscript{$\odot$}, while the SEP spread at 1 au is at least 222$^{\circ}$.
 
 \item[$\bullet$ Energetic particle propagation:] The differences in the SEP time profiles observed by MESSENGER (prompt and sharp) and STEREO-A (delayed and gradual) suggest that particles undergo significant interplanetary scattering between 0.3 and 1 au. The particle enhancement observed at L1 may be attributed to cross-field transport, and this is also the case for STEREO-B, at least until the spacecraft is  magnetically connected to the shock when it reaches $\sim$0.6 au, around the time the proton anisotropies presented an increase.   
 
\end{description}

 
 This work illustrates how a two-stage CME drives a single shock in the corona, which accelerates, at a relatively high heliocentric height and with no flare-related contribution, the energetic electrons and protons observed in this unusual SEP event. Regarding the observational differences between the SEP time profiles at MESSENGER (0.33 au) and STEREO-A (1 au), solar missions such as Parker Solar Probe \citep[][]{Fox2016}, launched in August 2018, or Solar Orbiter \citep[][]{Muller2020,Zouganelis2020}, launched in February 2020, will be crucial for a better understanding of the particle acceleration and transport in the inner heliosphere. The new era of combined in situ and remote-sensing multi-point observations has just started \citep[e.g.][]{Kollhoff2021}, using new data from different instruments like the Energetic Particle Detector \citep[EPD,][]{Rodriguez-Pacheco2020} on board Solar Orbiter.   
 
\begin{acknowledgements}
The authors thank an anonymous referee for a detailed review that improved the quality of the manuscript. The UAH team acknowledges the financial support by the Spanish Ministerio de Ciencia, Innovación y Universidades FEDER/MCIU/AEI Projects ESP2017-88436-R and PID2019-104863RB-I00/AEI/10.13039/501100011033. LRG is also supported by the European Space Agency, under the ESA/NPI program, and acknowledges the support from Anik De Groof, Andrew Walsh, Olivier Witasse, Georgina Graham, Ignacio Cernuda Cangas and Francisco Espinosa Lara. FCM acknowledges the financial support by the Spanish MINECO-FPI-2016 predoctoral grant. LB acknowledges the support from the NASA program NNH17ZDA001N-LWS (Awards Nr. 80NSSC19K1261 and 80NSSC19K1235). ND acknowledges financial support by DLR under grant 50OC1702. NVN's work has been supported by NASA contracts 80NSSC18K1126 and 80NSSC20K0287. MD acknowledges support by the Croatian Science Foundation under the project IP-2020-02-9893 (ICOHOSS). LKJ is supported by NASA’s Living with a Star and Heliophysics Supporting Research programs. The authors acknowledge the different Wind, ACE, SOHO STEREO, MESSENGER and Mars Odyssey instrument teams, and the STEREO and ACE science centers for providing the data used in this paper. In particular, we would like to thank Jean-Pierre Wuelser, Bill Thompson and Nathan Rich for their help regarding STEREO/SECCHI data. MESSENGER data were downloaded from the Planetary Data System. In addition, this paper uses data from the Heliospheric Shock Database, generated and maintained by the University of Helsinki, and from GONG, managed by the National Solar Observatory. The radio spectrograms used in this study are provided by the Observatoire de Paris-Meudon (\url{http://secchirh.obspm.fr/}). ENLIL simulation results have been provided by the CCMC at NASA Goddard Space Flight Center (GSFC) through their public Runs on Request system (\url{http://ccmc.gsfc.nasa.gov}; run IDs Laura\_Rodriguez-Garcia\_093020\_SH\_1, Laura\_Rodriguez-Garcia\_093020\_SH\_2). The WSA
model was developed by N. Arge, currently at GSFC, and the ENLIL Model was developed by D. Odstrcil, currently at George Mason University. 

\end{acknowledgements}

\begin{flushleft}

\textbf{ORCID iDs} 
\vspace{2mm}

Laura Rodríguez-García \orcid{https://orcid.org/0000-0003-2361-5510}

Raúl Gómez-Herrero \orcid{https://orcid.org/0000-0002-5705-9236}

Ioannis Zouganelis
\orcid{https://orcid.org/0000-0003-2672-9249}
Laura Balmaceda \orcid{https://orcid.org/0000-0003-1162-5498}

Teresa Nieves-Chinchilla
\orcid{https://orcid.org/0000-0003-0565-4890}

Nina Dresing \orcid{https://orcid.org/0000-0003-3903-4649}

Nariaki Nitta
\orcid{https://orcid.org/0000-0001-6119-0221}

Fernando Carcaboso Morales
\orcid{https://orcid.org/0000-0003-1758-6194}

Luiz Fernando Guedes dos Santos
\orcid{https://orcid.org/0000-0001-5190-442X}

Lan Jian \orcid{https://orcid.org/0000-0002-6849-5527}

Leila Mays \orcid{http://orcid.org/0000-0001-9177-8405}

David R Williams  \orcid{http://orcid.org/0000-0001-9922-8117}

Javier Rodríguez-Pacheco
\orcid{https://orcid.org/0000-0002-4240-1115}

\end{flushleft}
%
%

\end{document}